%% file: main.tex
\documentclass[11pt,a4paper]{article}
\usepackage{jinstpub}
\pdfoutput=1

\usepackage[colorinlistoftodos]{todonotes}
\usepackage[section]{placeins}
\usepackage[nottoc]{tocbibind}
\usepackage{caption}
\usepackage{subcaption}
\usepackage{lineno}
\linenumbers

\title{Performance of Michigan sMDT Prototype Chambers for the HL-LHC ATLAS Muon Detector Upgrade}

\author[a,1]{K. Nelson,\note{Corresponding author.}}
\author[a]{Y. Guo,}
\author[a]{D. Amidei,}
\author[a]{and E. Diehl}
\affiliation[a]{Department of Physics, University of Michigan,\\450 Church Street, Ann Arbor, MI 48109,  USA}

\emailAdd{kevin.nelson@cern.ch}

\abstract{A new small-diameter Monitored Drift Tube (sMDT) chamber has been developed for the muon spectrometer of the ATLAS experiment to handle the higher collision rates expected at the CERN High Luminosity Large Hadron Collider (HL-LHC).  This paper presents measurements of the tracking resolution and hit efficiency of two prototype sMDT chambers constructed at the University of Michigan. Using cosmic-ray muons the sMDT tracking resolution of 103.7$\pm8.1$ \textmu m was measured for one chamber and 101.8$\pm$7.8 \textmu m for the other, compared with a design resolution of 106 \textmu m. A further tracking resolution improvement to 83.4$\pm$7.8 \textmu m was obtained by using new high-gain readout electronics which will be added for HL-LHC.   An average tracking efficiency
of (98.5$\pm$0.2)\% was found for both chambers. The methodology used to determine the detector tracking resolution and efficiency, including reconstruction of sMDT data and a Geant4 simulation of the test chamber, is presented in detail. 
}


\begin{document}
\maketitle
\flushbottom
%
%
%
%
%
%
%
%
%
%

\section{Introduction}\label{sec:intro}
\input{Introduction}

\section{Methodology}\label{sec:method}
\input{Methodology}
\section{Multiple Coulomb Scattering}\label{sec:mc}
\input{MonteCarlo}
\section{Systematic Uncertainty Estimation}\label{sec:syst}
\input{Systematics}
\section{Results}\label{sec:results}
\input{Results}
\section{Conclusions}\label{sec:conclusions}
\input{Conclusions}

\acknowledgments

We would like to thank Bing Zhou, Curtis Weaverdyck, Xiangting Meng, Jing Li, and Zhe Yang
for construction of the prototype chamber and initial set up of the DAQ system for this study. We also thank Tiesheng Dai and Zhen Yan for their valuable comments and discussions during conducting these studies.  Finally, we thank the DOE for funding the ATLAS research project at Michigan (DOE Grant \#: DE-SC007859).

\bibliographystyle{JHEP}
\bibliography{bibliography}

\end{document}

%% file: Introduction.tex


The ATLAS muon spectrometer provides a muon trigger and measurement of muon momentum for the ATLAS experiment at the CERN Large Hadron Collider~\cite{AtlasCollaboration_2008}.  The spectrometer has cylindrical geometry and is 22 m in diameter and 45 m in length and covers $2\pi$ in $\phi$ and ±2.7 in pseudorapidity ($\eta$).  It
is divided into a barrel region and two endcaps each located inside an air-core toroidal magnet.  The magnetic field integral of the toroids ranges between 2.0 and 6.0 T m for most of the acceptance, and achieves a muon momentum resolution of 3\% at 100 GeV rising to 11\% at 1 TeV~\cite{TDRMuonSpectrometer}.  The barrel region
currently contains 3 concentric layers of Monitored Drift Tubes chambers (MDTs) which provide tracking with a single tube resolution of $80\mu$m, with 2 layers of Resistive Plate Chambers trigger chambers (RPCs) on the outer MDT layers.  The MDTs are arranged in 16 sectors covering $2\pi$ in $\phi$, in alternating "large" and "small" chamber sizes.  In the ATLAS Phase II upgrade, scheduled for 2024-25, an extra layer of small-gap RPC chambers will be added to the barrel inner layer to improve the efficiency and resolution of the barrel muon trigger~\cite{phase2TDRTDAQ}.  In order to accommodate the new RPCs in the upgrade, the "small" sector MDTs will be replaced with
small-diameter MDTs (sMDTs), which are are reduced in size compared to MDTs.  The "large" sectors already have space for the new RPCs, so the current MDTs will remain.

The sMDT chamber design was developed at the Munich Max Planck Institute  (MPI)~\cite{smdtDesign} and use drift tubes of 7.5 mm radius, compared to 15 mm for the original MDTs.  The smaller radius permits sMDTs to handle a $10\times$ higher hit rate compared to MDTs, and the smaller size creates space for the new RPC layers.  Aside from the difference in tube radius the ATLAS MDT and sMDT drift tubes are very similar. Both use aluminum tubes of 400 \textmu m  wall thickness, 50 \textmu m diameter W-Rh wires, and a drift gas admixture (Ar:CO$_2$, 93:7 \%) at 3 bar absolute pressure. To keep the gas gain ($2\times 10^4$) identical, the operating high voltage is 2730 V for sMDTs compared to 3080 V for MDTs.  These operating specifications were chosen in order for the nominal time-to-space transfer functions (r(t) functions) for the sMDT and MDT chambers to agree as closely as possible.  A summary of the design sMDT chambers with a  comparison MDTs can be found at \cite{smdtDesign}.

The University of Michigan and MPI Munich are sharing the production of sMDT chambers for the ATLAS Phase-II HL-LHC muon detector upgrade.  This study examines two prototype sMDT chambers built at the University of Michigan.

This paper describes the methodology for measuring the resolution and efficiency of sMDT chambers and present results of these measurements for two Michigan prototype sMDT chambers. These chambers must have high spatial resolution to allow an accurate momentum measurement using the sagitta of muon tracks in the ATLAS muon spectrometer.
The sMDT design single-hit resolution is 106  \textmu m~\cite{smdtDesign}.
Section \ref{sec:method} describes the experimental setup and the procedure for deriving the r(t) function for sMDT tubes and for fitting muon tracks.  Section \ref{sec:mc} describes the multiple scattering correction, and section \ref{sec:syst} describes the systematic uncertainties considered. Measured resolution and efficiency results are reported in section \ref{sec:results}.

%% file: Methodology.tex
\subsection{Prototype sMDT Chambers}

\begin{figure}
    \centering
    \includegraphics[width=0.5\textwidth]{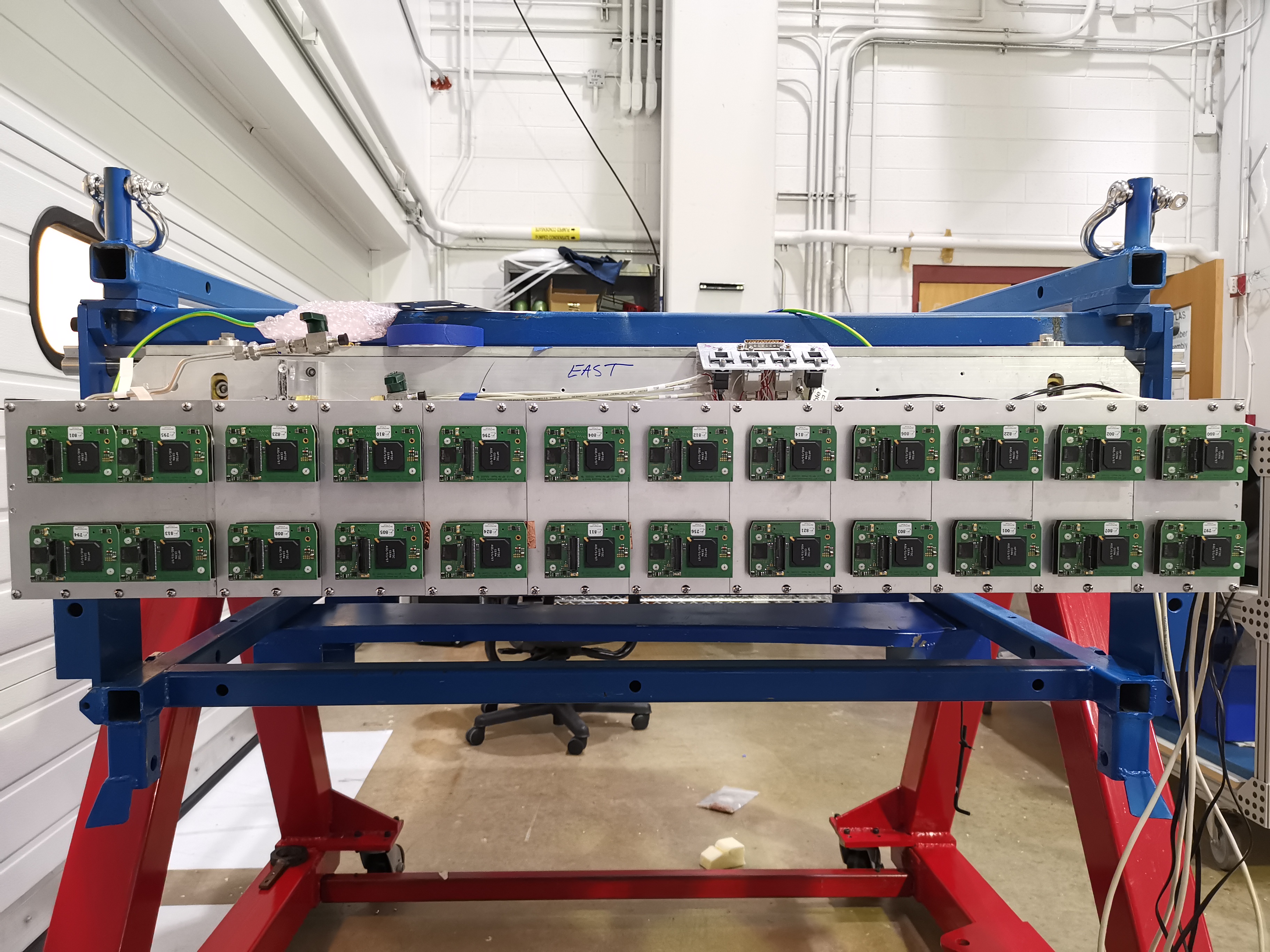}
    \caption{The BIS prototype chamber produced at the University of Michigan. The readout electronics are shown mounted on one end of the drift tubes. The other end of the tubes are fitted with high-voltage distribution boards. The chamber is approximately $0.2\times1.1\times1.5$ m$^{3}$ in size.}
    \label{fig:module0}
\end{figure}

The initial prototype chamber of the ATLAS BMG type was constructed in summer 2018 and a data acquisition (DAQ) system and tracking program were developed to test the prototype chamber.
In fall 2020, a prototype of the Phase-II upgrade sMDT BIS type chamber, shown in figure \ref{fig:module0}, was built to demonstrate the sMDT performance and to prepare for the sMDT mass production.  

The BMG and BIS prototype chambers both have 8 total tube layers, divided into two 4-layer multilayers which are separated by a spacer frame.  The horizontal and vertical wire pitch distances are identical. Table \ref{tab:smdt_proto_geo} summarizes the geometric properties of the two chambers. The tracking performance of the two chambers can be directly compared to confirm the robustness of the sMDT calibration and track fitting methods. Cosmic-ray muons were used to test the sMDT prototype chambers.

\begin{table}[]
    \centering
    \begin{tabular}{|l|c|c|} \hline
        Chamber & BMG & BIS \\ \hline 
        Layers & 8 & 8 \\ \hline 
        Tubes per layer & 54 & 70 \\ \hline
        Spacing between multilayers (cm) & 18 & 4.5 \\ \hline 
        Tube length (m) & 1.1 & 1.6 \\ \hline
        Tube wall thickness (mm) & 0.4 & 0.4 \\ \hline
    \end{tabular}
    \caption{Summary of geometric properties of the BMG and BIS prototype chambers.}
    \label{tab:smdt_proto_geo}
\end{table}

\subsection{Data Acquisition}\label{sec:DAQ}

\begin{figure}
    \centering
    \includegraphics[width=0.8\textwidth]{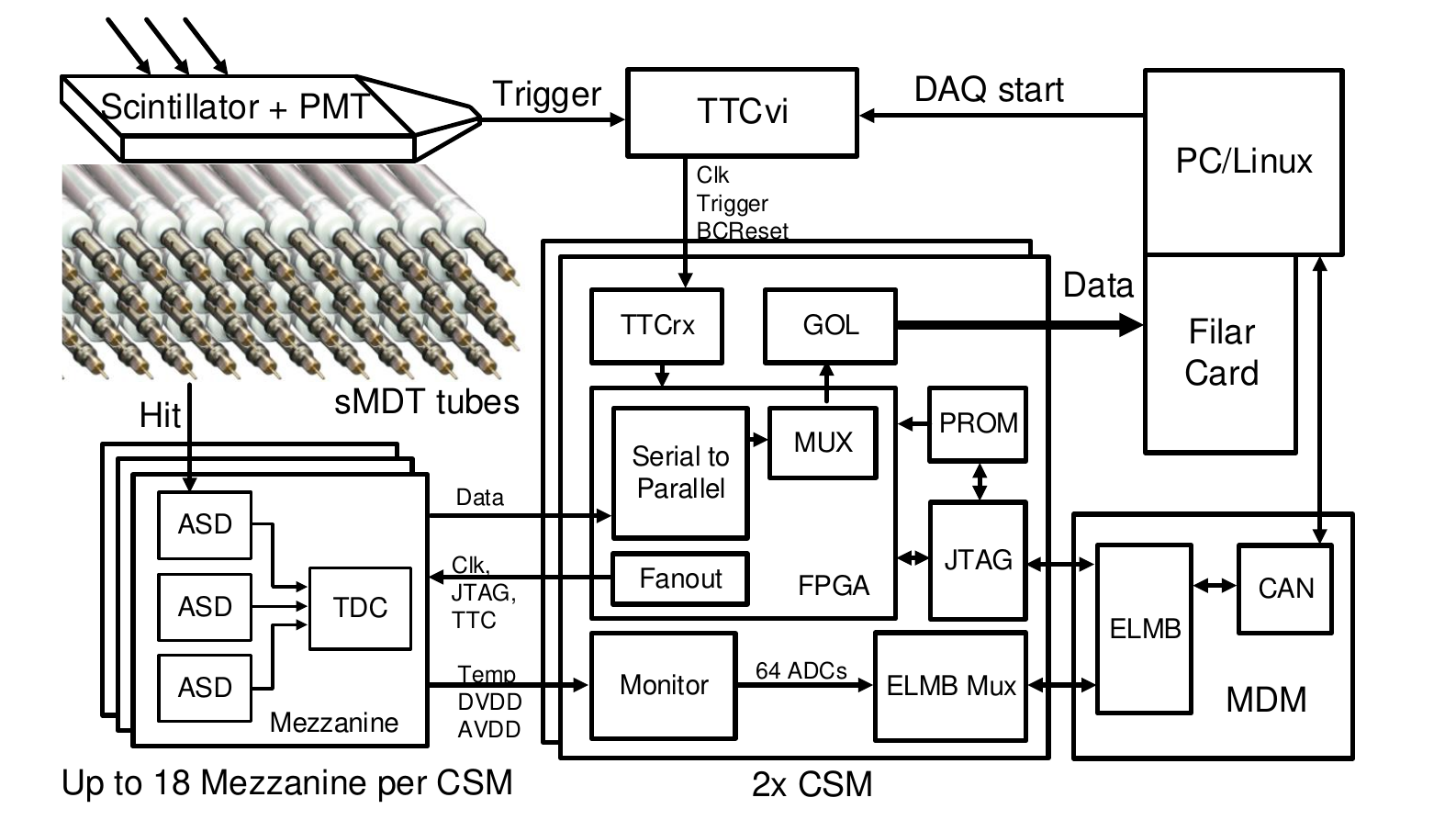}
    \caption{Block diagram of the sMDT readout electronics and mini-DAQ system. See text for explanation of components.}
    \label{fig:miniDAQ}
\end{figure}

\begin{figure}
    \centering
    \includegraphics[width=0.7\textwidth]{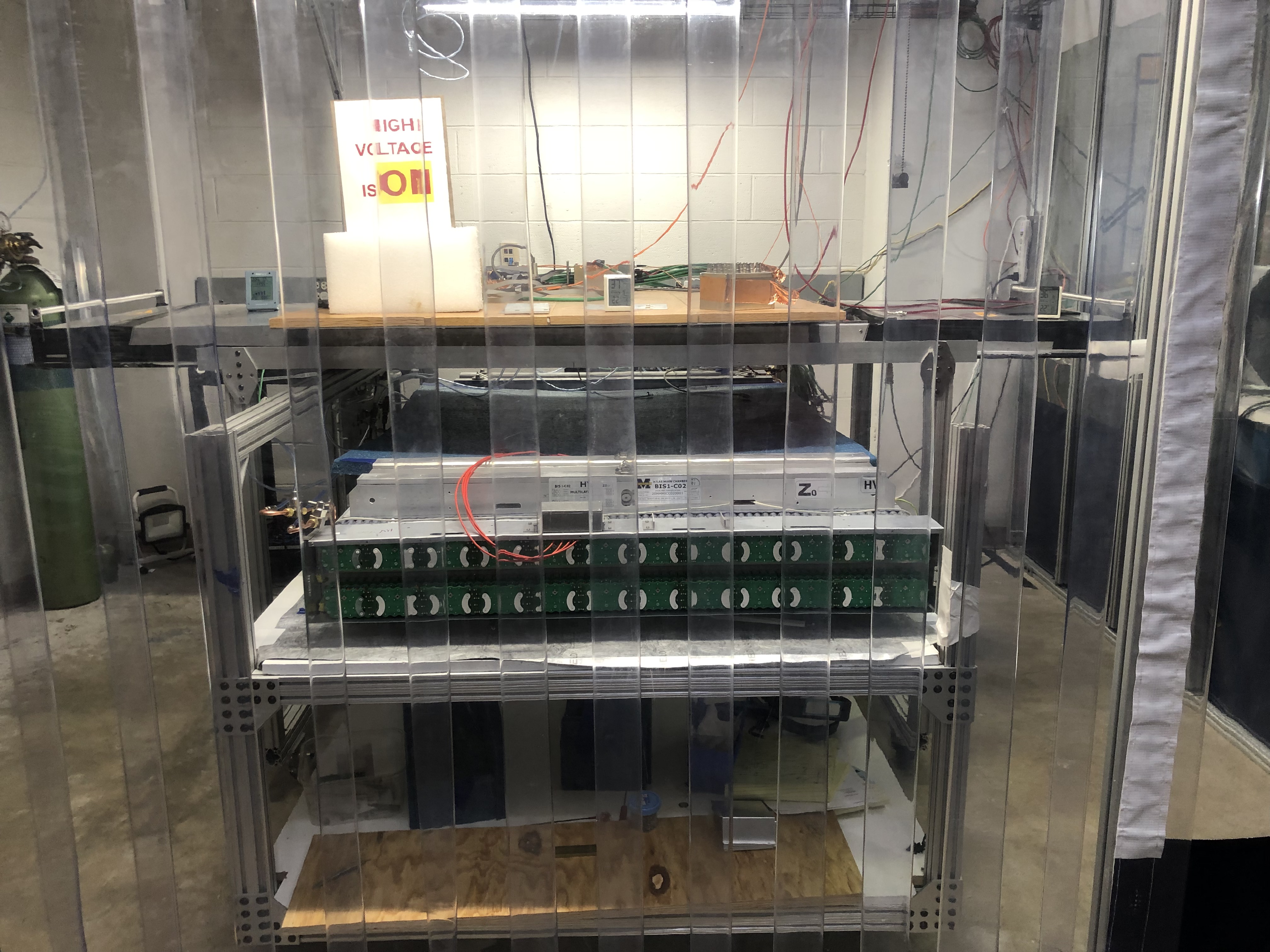}
    \caption{sMDT cosmic ray test station in a humidity-controlled room. The sMDT chamber sits on a cart with a $1.5 $m$^{2}$ scintillator mounted above.
    }
    \label{fig:CR-test}
\end{figure}

A mini-data acquisition (mini-DAQ) system \cite{TDCandMiniDAQ} developed at the University of Michigan, as shown in figure \ref{fig:miniDAQ}, is used to collect cosmic ray muon signals from the sMDT chamber. A large ($\approx1.5$ m$^2$) scintillator sits above the test chamber with photo-multiplier tubes (PMTs) mounted on each end and provides the trigger to the mini-DAQ by the coincidence of the two PMTs.  Front-end electronics is installed on 
a three-layer stack of circuit boards mounted on the end of the chamber.  The top circuit board is called the mezzanine card and contains the Amplifier/Shaper/Discriminator (ASD) \cite{asdManual} and the high performance time-to-digital converter (HPTDC) \cite{HPTDC} chips. The sMDT cosmic-ray test station is shown in figure \ref{fig:CR-test}.

When a muon passes through a tube, the electrons from the primary ionization clusters drift to the central wire which has a potential at 2730 V. The earliest arrival time of the ionized electrons to reach the wire is captured by the ASD and the drift time is obtained by comparing the difference between this arrival time and the scintillator trigger time. In addition, the pulse height of the detector signal for the first $\sim$20 ns is also measured by a Wilkinson ADC on the ASD, which allows the pulse-height-dependent slewing corrections to be adapted for the timing measurement. The pulse height of the signal is encoded as the time interval between the leading and trailing edges of the ASD output pulse. The HPTDC digitizes the both edges of the ASD output, performs the trigger matching, and sends the matched data to the Chamber Service Module (CSM) for readout. One CSM multiplexes data from up to 18 HPTDCs and transmits it to a Linux machine, where the data related to the same trigger is packaged as an event and saved for offline analysis. The scintillator trigger signal timing is also recorded by one channel of the HPTDC.

The TTCvi \cite{TTCvi} module is used to provide both the trigger and clock to the front end electronics. It receives the coincident PMT signal and then distributes this signal to the CSMs. An internal 40 MHz clock is also provided by the TTCvi to the CSMs. The CSMs then distribute the clock and trigger signal to all the mezzanine cards.

Triggering on cosmic-rays via scintillator PMTs has a significant trigger time smearing due to the large size of the scintillator, as will be addressed in section \ref{sec:analysis}. The ASD has a threshold of $\approx24$ primary ionization electrons corresponding to an electronics signal threshold of 39 mV. To filter out noise hits that are not related to a muon track, hits from adjacent tubes on different layers are grouped into clusters.  At least one but no more than three clusters of 1-10 hits is required in each multilayer of the chamber.  The requirement of at least one hit in each multilayer means some triggered events will not pass the reconstruction cuts.  The rejection of events with too many clusters vetos large shower events.  The predicted geometric acceptance is 21\% for data taken with ASD-1 and 73\% for data taken with ASD-2 (more ASD-2 cards were available so the instrumented proportion of the chamber increased).  The reconstruction cuts are approximately 30\% efficient on top of the geometric acceptance, and the majority of the events rejected with hits in both multilayers involve multiple and/or large clusters of hits, in which it appears a cosmic-ray particle has undergone a shower process, or coincident noise hits not adjacent to the primary cluster of hits.

Three different data-sets that have different front-end electronics and geometric coverage were used in this work. First, we took data for the prototype chamber with the legacy ASD (ASD-1) chips used for the MDT detector in ATLAS. 
Then, we collected data for BIS chamber with both the legacy ASD and the new, high gain ASD-2 chip which will be installed on all ATLAS MDT and sMDT chambers in the Phase-II upgrade.
When using legacy ASD chips, only a portion of the prototype chambers were instrumented due to limited availability of readout electronics. 
Therefore, only 24 of the 54 total tubes in each layer are read out using the data acquisition system. 

The results for the BIS prototype with ASD-2 use a fully instrumented chamber. The ASD-2 chip has approximately 50\% higher gain than ASD-1, which increases the precision on the timing measurement and reduces the size of the slew effect. 
Table \ref{tab:data_runs} summarizes the various data-taking configurations and numbers of events collected for each. Data with ASD-1 has significantly fewer events due to a limited availability of readout electronics. 
The cluster requirement was designed to reject events with coincident noise hits and select only the cleanest events. The figures in the remainder of the paper are from the prototype chamber with ASD-1, unless specified otherwise. The results reported in section \ref{sec:results} details the performance of all three data taking conditions.

\begin{table}[]
    \centering
    \begin{tabular}{|c|c|c|c|} \hline
        Chamber & ASD Version & Triggered Events & Reconstructed Events \\ \hline
        Prototype & ASD-1 & 30 million & 600,000 \\ \hline
        BIS & ASD-1 & 30 million & 1.9 million \\ \hline
        BIS & ASD-2 & 9.2 million & 2.0 million \\ \hline 
    \end{tabular}
    \caption{Summary of data-taking configuration for the data sets used in this analysis.  Triggered events is the total number of cosmic-ray events collected from the coincidence of the two PMTs.  
    Typical TDC and ADC spectra for each tube recorded in cosmic rays are shown in figure~\ref{fig:tdc_and_fit}.  Reconstructed events are those which pass the cluster requirement and which are used in r(t) calibration and resolution studies.}
    \label{tab:data_runs}
\end{table}

\begin{figure}[!htbp]
    \begin{subfigure}[b]{0.49\textwidth}
        \centering  
        \includegraphics[width=\textwidth]{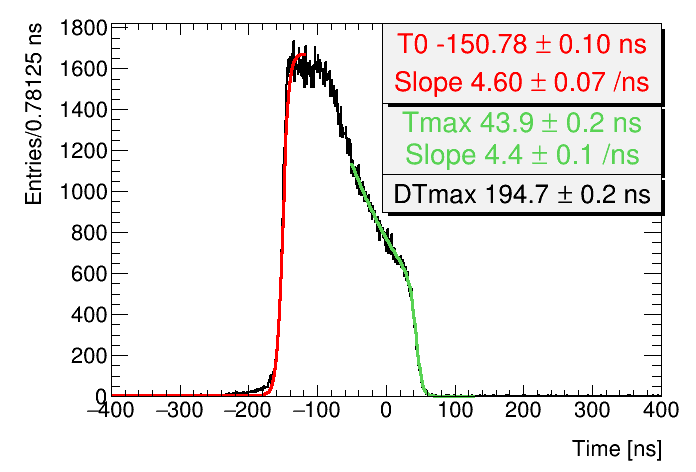}
        \caption{}
        \label{fig:tdc_fit}
    \end{subfigure}%
     \begin{subfigure}[b]{0.49\textwidth}
         \centering
         \includegraphics[width=\textwidth]{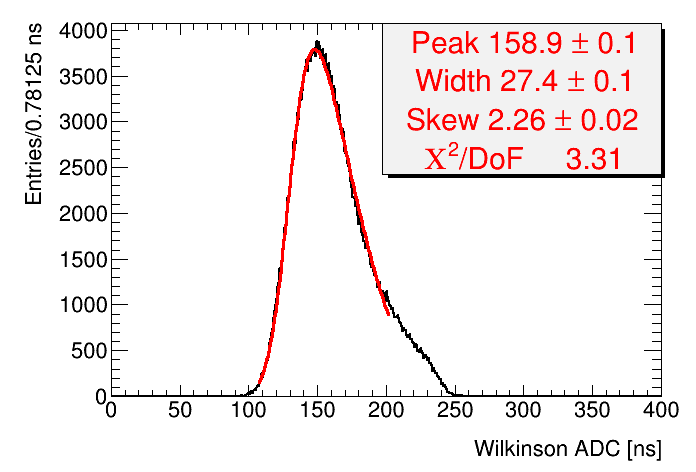}
         \caption{}
         \label{fig:adc_fit}
     \end{subfigure}%
    \centering
    \caption{(a) Example drift time spectrum for one channel from a cosmic-ray data run collected using the BMG prototype chamber. The red curve is the Fermi-Dirac fit of the rising edge to determine $t_0$.  The green curve is a fit the falling edge to determine $t_{max}$.  DTmax is $t_{max}-t_0$ (b) Example ADC spectrum for the same channel, where the red curve and text show the results of a skew normal fit.}
    \label{fig:tdc_and_fit}
\end{figure}

\subsection{Data Analysis}\label{sec:analysis}

The data analysis developed in this work encompasses two main areas: (1) calibrating the drift tubes and reconstructing tracks from raw data to determine the tracking resolution; (2)
estimating the multiple scattering of cosmic-rays so that this multiple scattering can be removed from the resolution measurement to allow comparison with ATLAS data from proton collisions. The analysis strategy is as follows. First, we correct the drift time for time slew and fit the rising and falling edges of the drift time spectrum to derive the minimum and maximum drift time for each tube. We then perform a calibration of the r(t) function, which is parameterized as a 10-degree Chebyshev polynomial.  We then fit straight, 2-D tracks in the plane perpendicular to the wires and create distributions of biased and unbiased tracking residuals.  The residual is the difference between the drift radius and the radius predicted by the straight line fit. Biased residuals are those from fits using all tube hits, whereas unbiased residuals are found by refitting a track multiple times by removing one hit from the from fits and finding the residual of the hit removed from the fit.  To account for multiple coulomb scattering, which is much more significant for cosmic-rays than muons in the ATLAS experiment with $\geq$ 20 GeV of $p_T$, residual distributions are deconvoluted with a Monte Carlo (MC) truth multiple scattering residual distribution.\footnote{$p_T\geq$ 20 GeV is the cut used in the ATLAS MDT Run 2 resolution measurement \cite{MDTResolution}.}  Finally, the spatial resolution is calculated.

We now describe the procedure in greater detail. A time slew correction is subtracted from the drift time to compensate for the timing jitter caused by variations in the signal amplitude.  Large signals have shorter rise time and cross the ASD threshold more quickly than small signals causing a pulse-size dependent time slew. The time slew is a function of the response of the particular Amplifier/Shaper/Discriminator (ASD) chip used in the front-end electronics, as described in the corresponding manual (see figures 34 and 35 in \cite{asdManual}).  The time slew correction used in this study is taken from ATLAS MDT calibrations and shown in equation  \ref{eq:timeslew}: 

\begin{equation}\label{eq:timeslew}
    \text{timeslew}=35.59e^{-ADC/61.33\text{ns}} ~\text{[ns]}
\end{equation}

where ADC is the output pulse width of the ASD chip.  Typical values for ADC are on the order of 100 nanoseconds, and typical values for the time correction are $\approx6\pm2.5$ ns.  The nonzero mean slew correction is irrelevant and will be accounted for in the $t_0$ calibration described below.  

The signals also contain an timing offset, $t_0$, due to electronics and other delays, which must also be subtracted from the drift time.  The $t_0$ is defined as the half-way point in the leading edge of the drift time spectrum.   The  $t_0$ is determined by fitting the leading edge of the drift time spectrum with a Fermi-Dirac step function as defined in equation \ref{eq:tmin} and shown in figure \ref{fig:tdc_fit}:   

\begin{equation}\label{eq:tmin}
    b + \frac{A}{1+e^{-(t-t_0)/T}}
\end{equation}

where $b$ is the background noise floor, $A$ is the amplitude, $T$ is the $t_0$ slope (rise time).  In addition, the maximum drift time, $t_\text{max}$, is determined by doing a Fermi-Dirac fit on the falling edge of the drift time spectrum.

The ADC distribution is fit using a skew normal distribution.  Note that in order to cut noise, a threshold cut of 40 ns was imposed on ADC values.
Example drift time and ADC spectra are shown in figure \ref{fig:adc_fit}.  




Auto-calibration~\cite{autocalibration} and track fitting are both performed using a linearized least-squares procedure.  Auto-calibration is the process by which the parameterization describing the r(t) function is updated iteratively.  The auto-calibration algorithm iterates over two steps: (i) fitting tracks for set of events using a fixed r(t) function; (ii) updating the r(t) function using the least-squares method.  The process continues until the parameterization of the r(t) function converges.  Auto-calibration requires knowledge of the resolution as a function of drift radius, which is the quantity we ultimately wish to measure.  For the r(t) auto-calibration
we initialize the resolution function to that observed using ATLAS MDTs \cite{MDTResolution}.

The r(t) function is parameterized as a Chebyshev polynomial with 10 terms.  Via the least-squares method we calculate the changes to apply to each Chebyshev coefficient on each iteration of auto-calibration.  Chebyshev functions are defined on the domain [-1,1], so we linearly scale the domain of the Chebyshev functions to [$t_0$, $t_\text{max}$] for the individual $t_0$, $t_\text{max}$ for each tube.  We also constrain the r(t) function so that the predicted radius r($t_0$)=0.  Without this constraint the auto-calibration will tend to fit an unphysical nonzero radius for the minimum drift time.

\begin{figure}[!htbp]
    \centering
    \includegraphics[width=0.5\textwidth]{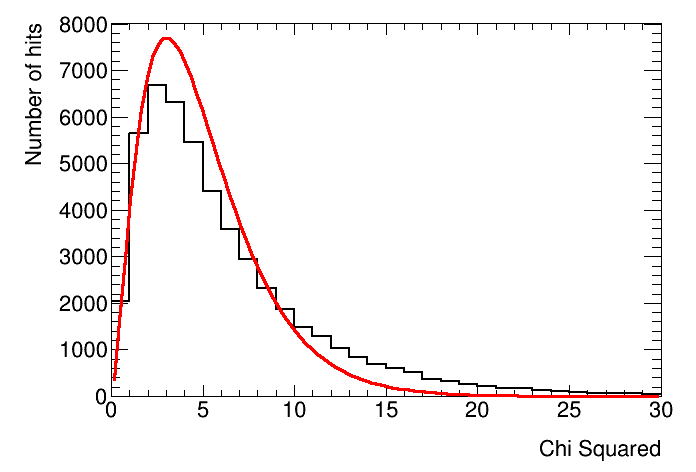}
    \caption{Chi-square distribution for fitted straight line tracks.  The tracks are fit in a 2-D plane perpendicular to the wires, as shown in figure \ref{fig:eventDisplay}.  The red curve shows the expected chi-square distribution, given the number of degrees of freedom, normalized to the same area as the histogram.}
    \label{fig:chi2}
\end{figure}

\begin{figure}[!htbp]
    \centering
    \includegraphics[width=0.5\textwidth]{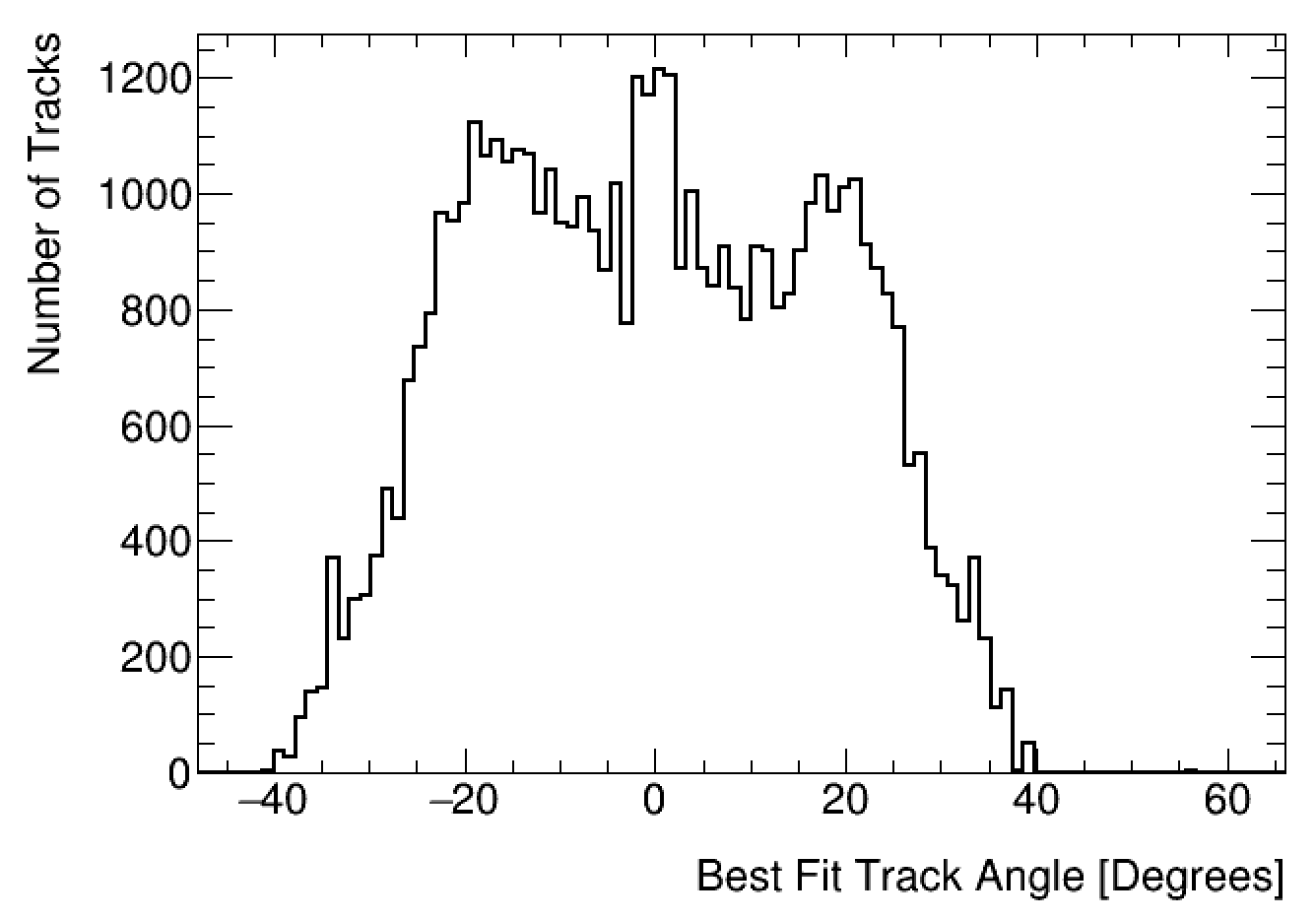}
    \caption{The angular distribution for all fitted tracks for 24 hour cosmic-ray run taken with the BMG prototype.}
    \label{fig:angulardist}
\end{figure}

\begin{figure}[!htbp]
    \centering
    \includegraphics[width=0.5\textwidth]{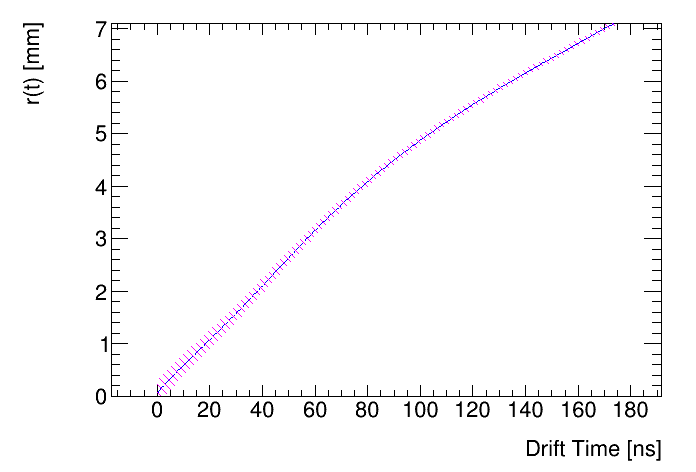}
    \caption{The auto-calibrated r(t) function.  Note that the Chebyshev polynomials are defined on the range [-1,1] so we linearly scale this r(t) function to [$t_0$,$t_\text{max}$] for each tube.  Therefore, different tubes may have slightly different r(t) functions if they have a differing maximum drift time $t_\text{max}-t_0$.  The hatched region represents the resolution as a function of radius presented in figure \ref{fig:resVsRadius}.}
    \label{fig:rtfunction}
\end{figure}

A robust means of validating this procedure is to compare the chi-square distribution obtained to the expected one with the same number of degrees of freedom.  The chi-square distribution obtained from the track fits using the auto-calibrated r(t) function is shown in figure \ref{fig:chi2}, compared to the predicted chi-square distribution.  The non-closure in the tail is likely the result of multiple coulomb scattering, which is not accounted for at this stage in the analysis, but will be addressed in section \ref{sec:deconvolution}.  The angular distribution for all fitted tracks is shown in figure \ref{fig:angulardist}. The r(t) function for the prototype chamber is shown in figure \ref{fig:rtfunction}.  The error band is the resolution as a function of radius, shown in figure \ref{fig:resVsRadius}.

Track fitting is done as follows. The first step is to make an initial guess of track parameters via a simple pattern recognition algorithm to seed the chi-square minimization algorithm.
The pattern recognition algorithm used in this work is a brute force method to determine if the track passes to the right or left of the wire.  We try $2^\text{(n hits)}$ different combinations of hit location, displaced either to the right or left of the wire by the radial distance.  The combination of right/left decisions that has the best chi-square is used to provide initial values of the slope and intercept of the track.  For the first pass, the error used is the expected MDT resolution function \cite{MDTResolution}.  After the radial resolution function is measured, we iterate the entire resolution procedure until the coefficients in table \ref{tab:resVsRadiusPolyPars} converge.

\begin{figure}[!htbp]
    \centering
    \includegraphics[width=0.5\textwidth]{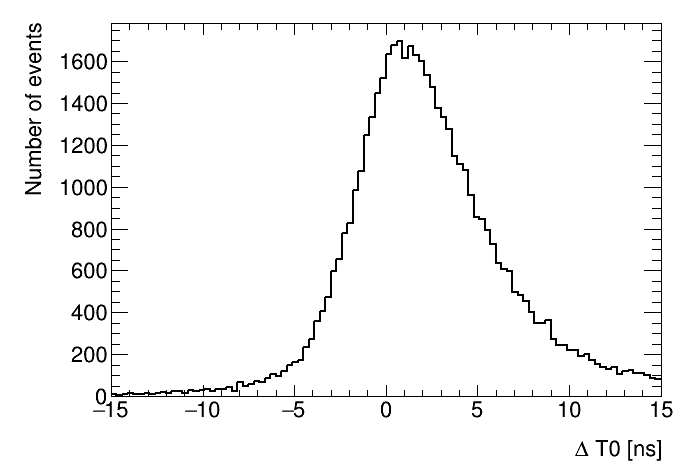}
    \caption{$\Delta t_0$ distribution from fit residual calculation for cosmic-ray data taken with the BMG prototype.
    }
    \label{fig:deltaT0}
\end{figure}

\begin{figure}[!htbp]
    \centering
    \includegraphics[width=0.5\textwidth]{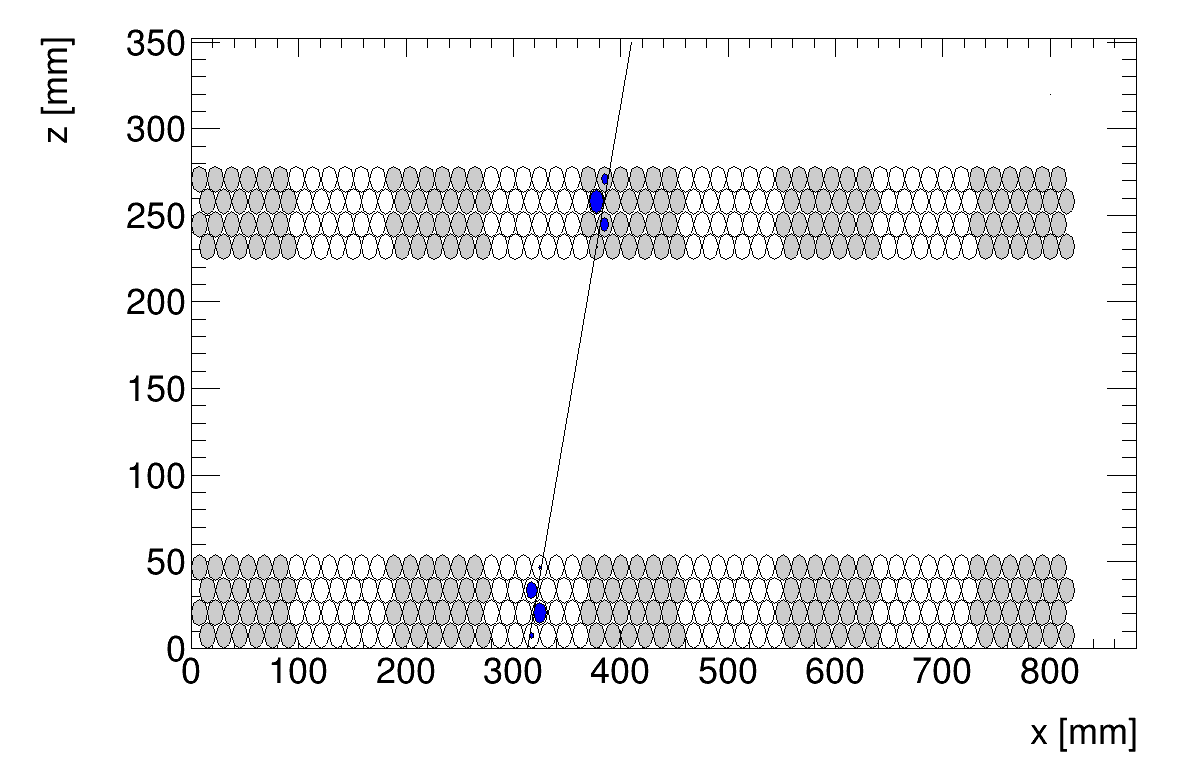}
    \caption{A single cosmic ray muon event in the BMG prototype chamber with a straight line track fitted.  The blue circles show the drift radius of each hit, which is determined using the auto-calibrated r(t) function.  The grey and white regions show the demarcations between different readout electronics cards.}
    \label{fig:eventDisplay}
\end{figure}

Track fitting again uses the linearized least-squares technique.  Tracks are modeled in a 2-dimensional plane and can be parameterized as a straight line with an angle with respect to the vertical and x-intercept (see figure \ref{fig:eventDisplay} for a definition of the x axis).
The choice of angle with respect to the vertical and x-intercept to describe a straight line avoids possible singularities in slope and y intercept for a purely vertical track.   Additionally, tracks have a third parameter: a global time shift $\Delta t_0$.   The $\Delta t_0$ is introduced to account for the random variations in trigger time from cosmic rays which are unsynchronized with the clock used by the DAQ. The $\Delta t_0$ distribution shown in figure \ref{fig:deltaT0}.

\begin{figure}
\begin{subfigure}[b]{0.49\textwidth}
        \centering  
        \includegraphics[width=\textwidth]{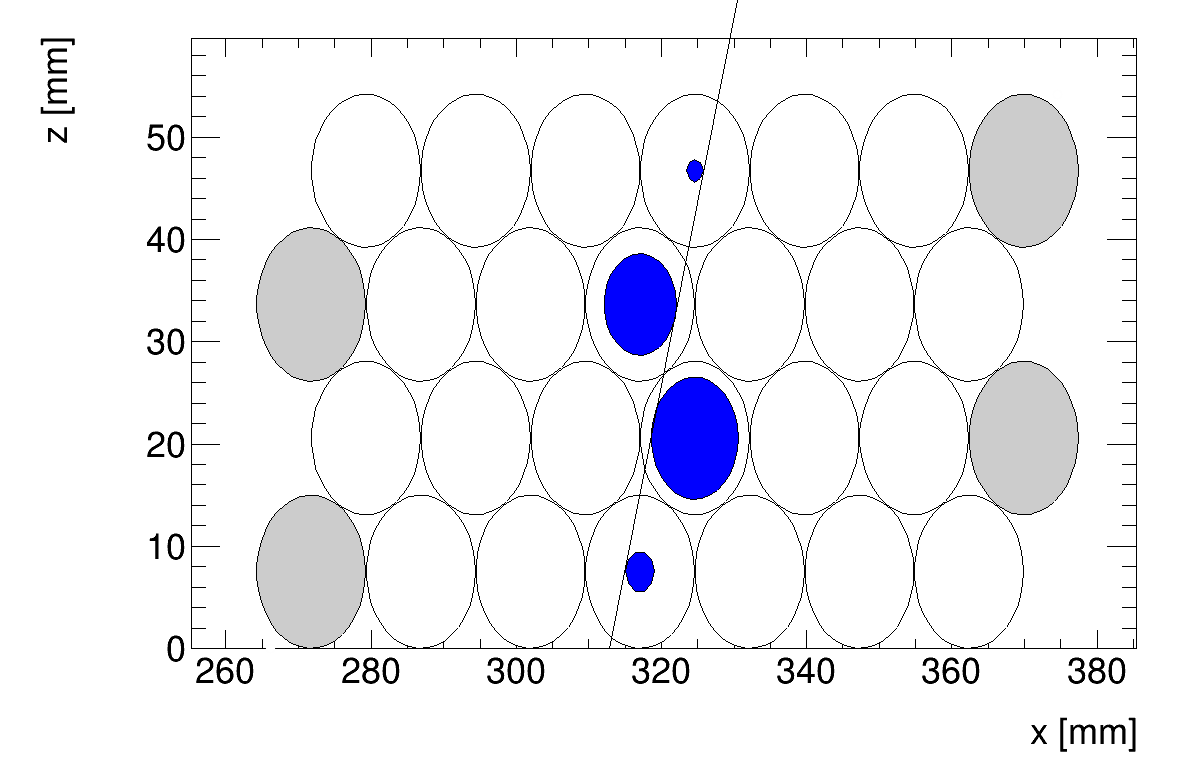}
        \caption{}
        \label{fig:ed_ml0}
    \end{subfigure}%
     \begin{subfigure}[b]{0.49\textwidth}
         \centering
         \includegraphics[width=\textwidth]{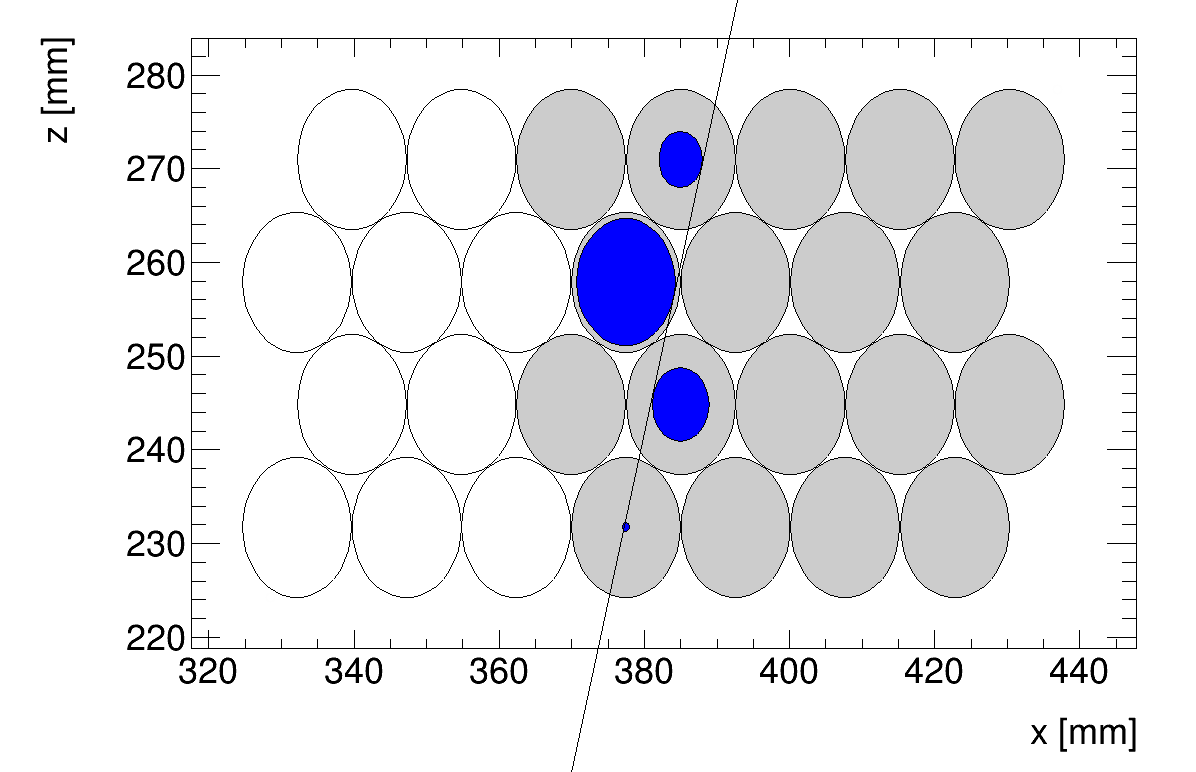}
         \caption{}
         \label{fig:ed_ml1}
     \end{subfigure}%
    \centering
    \caption{Close up of the (a) bottom and (b) top multilayers for the event shown in figure \ref{fig:eventDisplay}.  The blue circles show the radius of each hit, which is determined using the auto-calibrated r(t) function.  The grey and white regions show the demarcations between different readout electronics cards.  The black line is the track fit.}
    \label{fig:eventDisplayZoom}
\end{figure}

Tracks are fit using the current r(t) function, and residuals from the best fit track are used in the auto-calibration routine to update the Chebyshev coefficients.  An example event display with a fitted two-dimensional track is shown in Figures \ref{fig:eventDisplay} and \ref{fig:eventDisplayZoom}. 

\subsection{Resolution}\label{sec:resolution}

After calibrating the r(t) function we performed a resolution measurement.  Using the calibrated r(t) we fit a straight line track in the two dimensional plane perpendicular to the wires, with the additional event-by-event $\Delta t_0$ timing shift and find the biased and unbiased residual distributions.   We define the quantities $\sigma_b$ (biased width) and $\sigma_u$ (unbiased width) as the average width of a double Gaussian fit to the corresponding residual distribution, weighted by the amplitude of each Gaussian \cite{MDTResolution}:

\begin{equation}
    \sigma_{[b, u]}=\frac{A_w\times\sigma_w + A_n\times\sigma_n}{A_w+A_n}
    \label{eq:weightedsigma}
\end{equation}

\begin{figure}
\begin{subfigure}[b]{0.49\textwidth}
        \centering  
        \includegraphics[width=\textwidth]{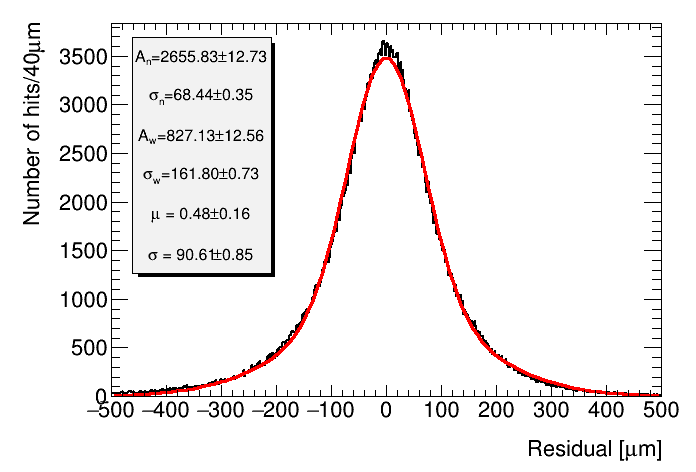}
        \caption{}
        \label{fig:fitResiduals}
    \end{subfigure}%
     \begin{subfigure}[b]{0.49\textwidth}
         \centering
         \includegraphics[width=\textwidth]{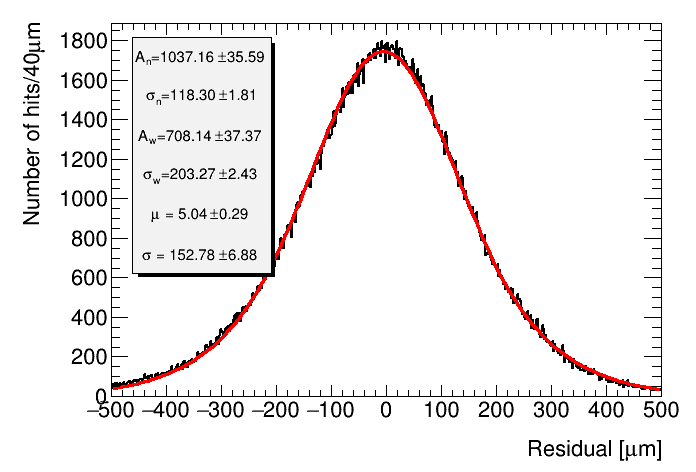}
         \caption{}
         \label{fig:hitResiduals}
     \end{subfigure}%
    \centering
    \caption{(a) Biased residuals from a cosmic-ray run taken with the BMG prototype. The red curve shows the double Gaussian fit. The fit parameters shown in the box are: $A_{n}$, $\sigma_{n}$ = Amplitude and width of the narrow Gaussian fit, respectively; $A_{w}$, $\sigma_{w}$ = Amplitude and width of the wide Gaussian fit, respectively; $\mu =$ common mean of wide and narrow fits;  $\sigma = $ the biased residual distribution width from eq. \ref{eq:weightedsigma}. (b) Unbiased residuals from cosmic-ray data taken with the BMG chamber.  The red curve shows the double Gaussian fit.  The fit parameters are the same as shown in (a) except $\sigma = $ the unbiased residual distribution width using eq. \ref{eq:weightedsigma}.}
    \label{fig:biasedAndUnbiasedResiduals}
\end{figure}

Where $A$ is the amplitude of the Gaussian, $\sigma$ is its standard deviation, and the subscripts $n$ and $w$ are the narrow and wide Gaussians of the double Gaussian distribution.  The double Gaussian fit is constrained to have a common mean for the two Gaussians.  
Examples of biased and unbiased residual distributions with double Gaussian fits are shown in figure \ref{fig:fitResiduals} and figure \ref{fig:hitResiduals}, respectively.  The resolution $\sigma$ is defined to be the geometric mean of the two residual distributions \cite{Carnegie}:

\begin{equation}\label{eq:resolution}
    \sigma = \sqrt{\sigma_b\times \sigma_u}
\end{equation}

Applying equation \eqref{eq:resolution} to the biased and unbiased residuals from the cosmic-ray data obtained using the BMG prototype chamber,
we find a single-hit resolution of 117.7 $\pm 2.1$ \textmu m. The quoted error is the statistical error from the fit parameters, propagated through equation \eqref{eq:resolution}.  The primary reason for the tension with the expected resolution of 106 \textmu m is multiple scattering of low energy cosmic-rays, as will be discussed in section \ref{sec:mc}.

\subsection{Efficiency}\label{sec:efficiency}

We define two types of efficiency for the drift tubes: layer and tube.  The layer efficiency is defined as the number of times a track passes through a layer with a hit recorded divided by the number of times a track passes through a layer regardless of whether a hit was recorded. 
The layer efficiency accounts for both the efficiency of the tubes in reconstructing hits and the dead regions due to tube walls and the gaps between tubes.  

Tube efficiency is the number of tracks passing through the gas volume of a tube with a hit recorded divided by the number of tracks passing through the gas volume of that tube whether or not a hit was recorded.

We perform the entire reconstruction methodology using the relaxed cuts just described: decoding raw data, fitting the $t_0$ and $t_\text{max}$ of drift time distributions, auto-calibration, and track fitting.  We then calculate the number of times the best fit track went through the layer (or tube) and did or did not register a hit.  When evaluating whether a track passes through a tube volume we leave that tube out of the track fit so as not bias the fit to be closer to that tube, as is done to calculate the unbiased residual distribution.  Hits greater than 5$\sigma$ away from a track are not counted towards the efficiency.

%% file: MonteCarlo.tex
\subsection{Geant4 Simulation}

\begin{figure}[!htbp]
    \centering
    \includegraphics[width=0.5\textwidth]{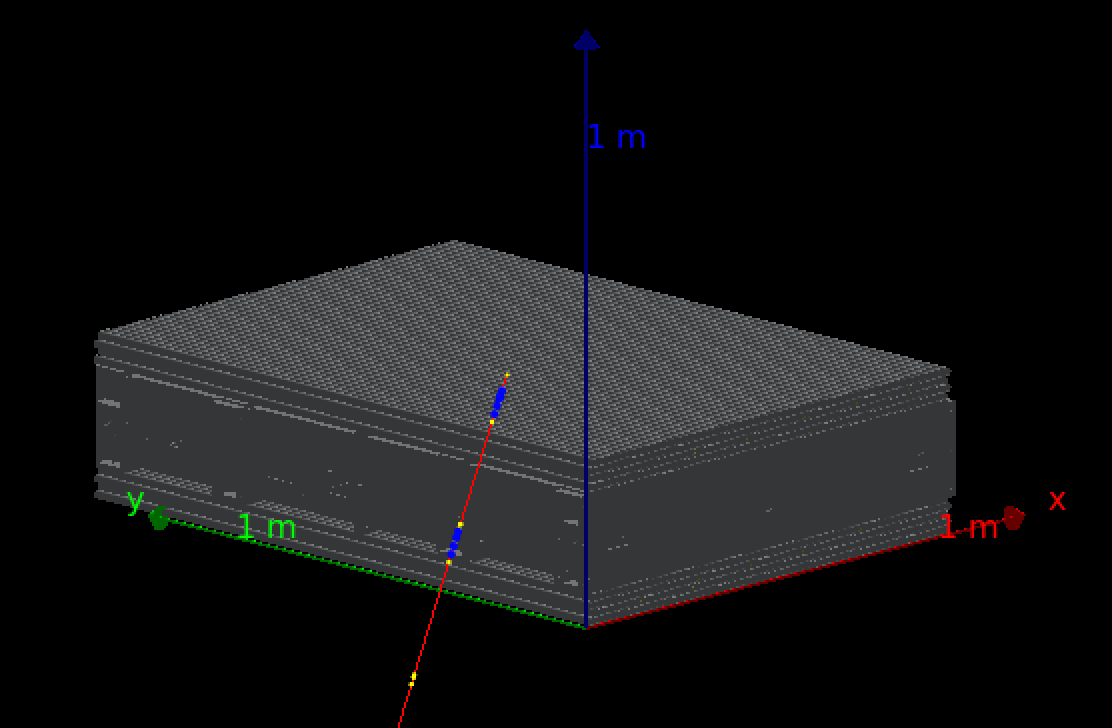}
    \caption{A 1 GeV muon passes through the simulated BMG prototype chamber. The blue markers show where the muon passed through a drift tube and the yellow markers show where the muon scatters from the air or the chamber.}
    \label{fig:chamberGeo_angle}
\end{figure}

In order to estimate the impact of multiple scattering, we created a Geant4 \cite{Geant4} simulation of the BMG prototype chamber.  The simulated geometry includes the tubes (walls, gas, and wires) and the spacer frame, which sits between the multilayers.  The spacer frame represents a significant amount of material, and has a complex geometry.  
Figure \ref{fig:chamberGeo_angle} shows a 1 GeV muon passing through the simulated chamber in Geant4.

\begin{figure}
    \centering
    \includegraphics[width=0.5\textwidth]{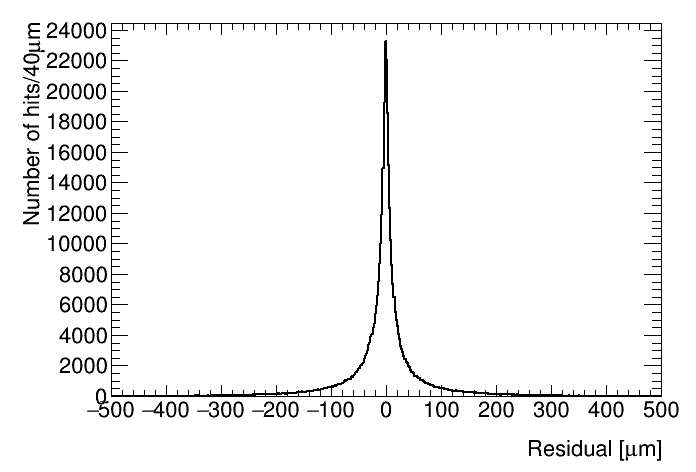}
    \caption{Residual distribution obtained from simulating multiple scattering of cosmic-rays passing through the test chamber.  There is a sharp central peak and non-Gaussian tails.}
    \label{fig:MC_residuals}
\end{figure}

Using MC truth information for charged particles passing through the simulated gas volume, we reconstruct events, similarly to what was done from raw data.  Thus, we can create a dataset with perfectly reconstructed hit radii.  In practice the hit radii are reconstructed as the closest approach of any charged particle within the gas volume, so the effects of delta rays or any other energetic secondaries are included.  This data is fed through exactly the same data analysis steps discussed in section \ref{sec:analysis} and section \ref{sec:resolution}.  The width of the residual distributions are solely the result of multiple scattering, since tube resolution has not been added to the monte carlo.  For an example residual distribution see figure \ref{fig:MC_residuals}.  Unlike the residual distribution observed in data, it is not well described by a double Gaussian fit, but has a sharp central peak and non-Gaussian tails.

As validation of the r(t) calibration procedure, we have run on an MC truth dataset with 20 GeV muons which have very little multiple scattering, with hit radii reconstructed as described.  Hit times are then created with the inverse r(t) function,
which is linearly interpolated in 1.9 ns bins.  In such a test over 90\% of the fit residuals are less than 10 microns, validating the ability of the auto-calibration to capture a realistic r(t) function.

\subsection{Simulation of the Cosmic-Ray Spectrum}

The effect of the multiple scattering is dependent on the assumptions on the particle and energy content of the cosmic-ray spectrum.  Many particle species are present in cosmic-rays due to energetic electromagnetic and hadronic showers in the upper atmosphere.  In this work we simulated only the muon and soft-electron components of the cosmic-ray spectrum, which are the most significant components at sea level.

\begin{figure}[!htbp]
    \centering
    \includegraphics[width=0.5\textwidth]{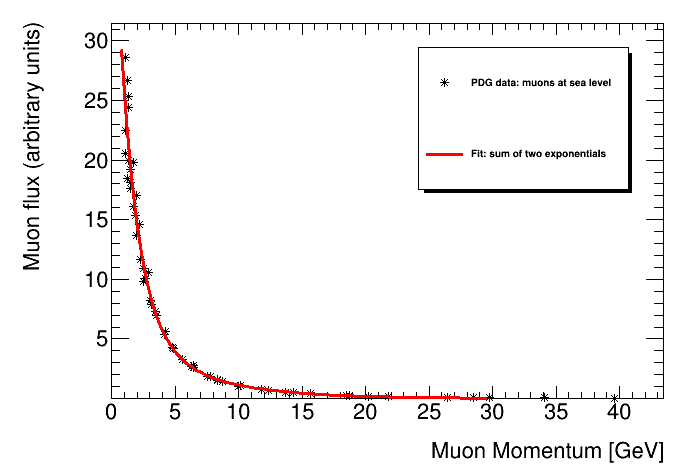}
    \caption{Probability distribution for muon flux at sea level.  The sum of two exponential curves are fit to cosmic-ray data.  For functional form, see equation \eqref{eq:muonFlux}.  For fit parameters, see table \ref{tab:fluxParameters}.}
    \label{fig:muonFlux}
\end{figure}

The Particle Data Group (PDG) article on cosmic-rays \cite{PDG} provides data describing muon flux as a function of momentum at sea level. We found this data was well fit as the sum of two exponentials as seen in figure \ref{fig:muonFlux}. The functional form of the probability distribution is described in equation \eqref{eq:muonFlux}, and the coefficients are listed in table \ref{tab:fluxParameters}.  The PDG article also notes that the muon energy spectrum is roughly flat below 1 GeV, so the fit terminates at 0.6 GeV, where the data runs out, and we simulate the energy spectrum as flat below this cutoff.

\begin{equation}\label{eq:muonFlux}
    \frac{A_1}{\lambda_1} e^{p/\lambda_1} + \frac{(1-A_1)}{ \lambda_2} e^{p/\lambda_2}
\end{equation}

\begin{table}[]
    \centering
    \begin{tabular}{|c|c|} \hline
        Parameter & Value \\ \hline 
        $A_1$ & 0.3924 \\
        $\lambda_1$ & 5.5264 \\
        $\lambda_2$ & 1.4180 \\ \hline
    \end{tabular}
    \caption{Fitted parameters to equation \eqref{eq:muonFlux} to describe the muon component of the cosmic-ray spectrum}
    \label{tab:fluxParameters}
\end{table}

The muon component described thus far is estimated to be 70\% of the particle content.  The remaining 30\% of cosmic-rays are simulated as soft-electrons.  This ratio is taken from the PDG article on cosmic-rays~\cite{PDG}, which lists the intensity of cosmic-ray muons to be $\approx166 \text{m}^{-2} \text{s}^{-1}$ and $\approx55 \text{m}^{-2} \text{s}^{-1}$ for cosmic-ray electrons above 10 MeV, for a $\approx$25\% electron component (integrated within our geometric acceptance).  Therefore 30\% is a conservative estimate of the electron component of the cosmic-ray spectrum.
Ultimately, we will show in section \ref{sec:softEsyst} that the resolution is rather insensitive to the soft-electron fraction.  The electron component is modeled as having a minimum energy of 10 MeV and as an exponential with a length scale of 50 MeV.

\subsection{Deconvolution of Multiple Scattering Effects}\label{sec:deconvolution}

The observed residuals presented in section \ref{sec:resolution} are a convolution of the true detector resolution and multiple scattering.  Using the MC truth multiple scattering residual distribution, we can remove the effect of multiple scattering via deconvolution and estimate the resolution of this detector for high $p_T$ muons in the ATLAS experiment.

\begin{figure}[!htbp]
    \centering
    \includegraphics[width=0.5\textwidth]{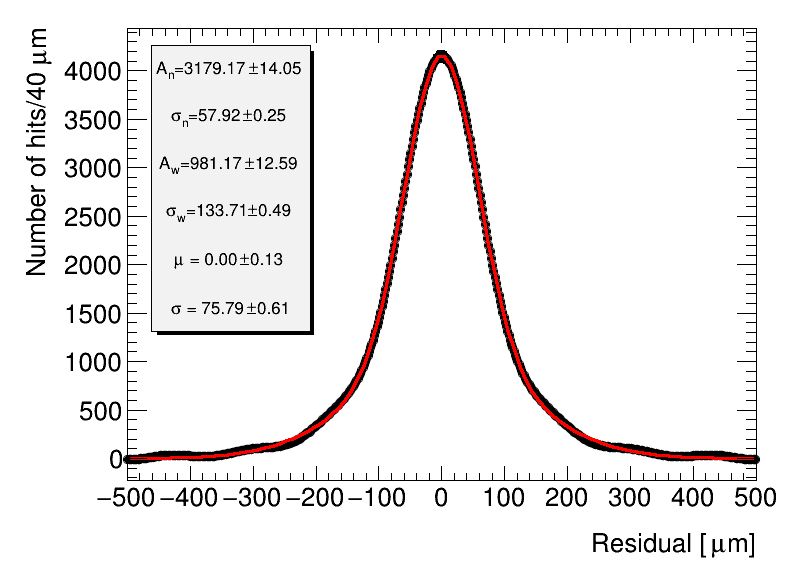}
    \caption{Biased residual distribution of the BMG chamber  from figure \ref{fig:fitResiduals} deconvoluted with the simulated multiple scatter residual distribution in figure \ref{fig:MC_residuals}. The red curve is a double Gaussian fit with fit parameters as described in figure \ref{fig:fitResiduals} shown in box.}
    \label{fig:deconvolution}
\end{figure}

The deconvolution is performed using Fourier transforms of the two input distributions.  In frequency space, a deconvolution is simply the division of the two histograms.  We Fourier transform the observed residuals, divide by the Fourier transform of the MC truth multiple scattering distribution, and then take the inverse Fourier transform of the result.  This result of the deconvolution procedure is shown in figure \ref{fig:deconvolution}.

We then perform a similar procedure to the deconvolution using a convolution to smear the residuals with the multiple scattering distribution of a sample of pure 20 GeV muons from the MC simulation.  20 GeV muons are chosen to be in accordance with the Run 2 MDT resolution result \cite{MDTResolution}, which placed a 20 GeV $p_T$ cut on muons.  Muons with 20 GeV of $p_T$ have at least 20 GeV of energy, so the 20 GeV monochromatic sample we used was conservative compared to the actual spectrum used in the Run 2 MDT resolution measurement.  The multiple scattering is much less for this sample, but  still adds 3-4 microns to the overall resolution.

%% file: Systematics.tex
\subsection{Auto-calibration and r(t) parameterization}

\begin{figure}[!htbp]
    \centering
    \begin{subfigure}[b]{0.49\textwidth}
         \centering
         \includegraphics[width=\textwidth]{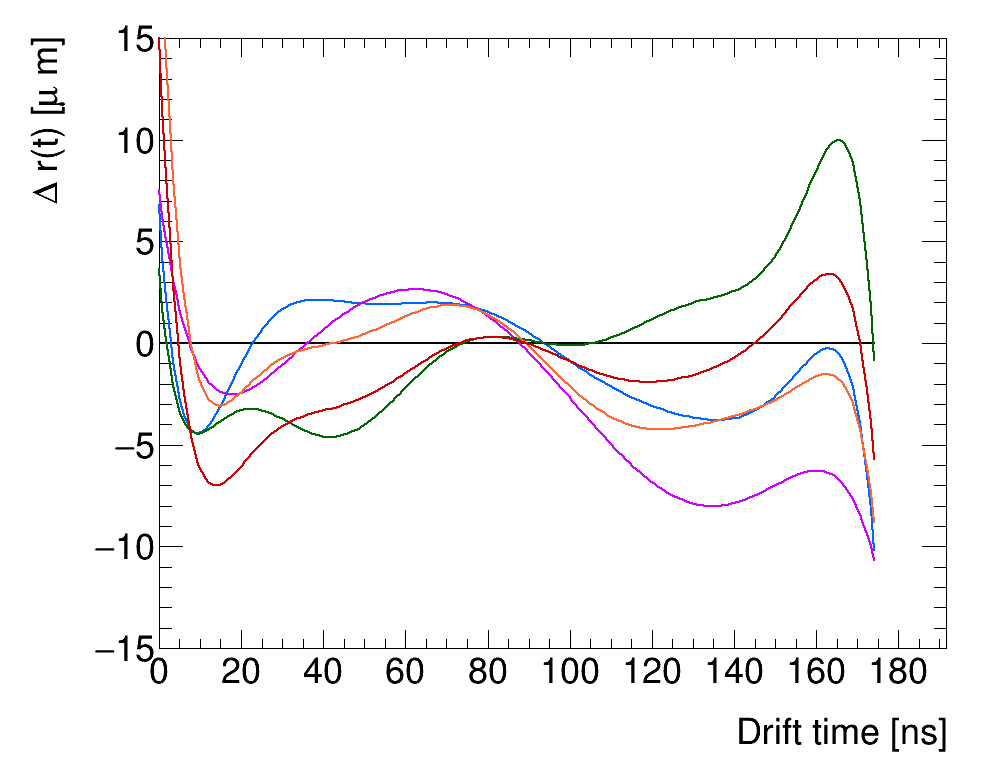}
         \caption{}
         \label{fig:manyRT}
     \end{subfigure}
     \hfill
     \begin{subfigure}[b]{0.49\textwidth}
         \centering
         \includegraphics[width=\textwidth]{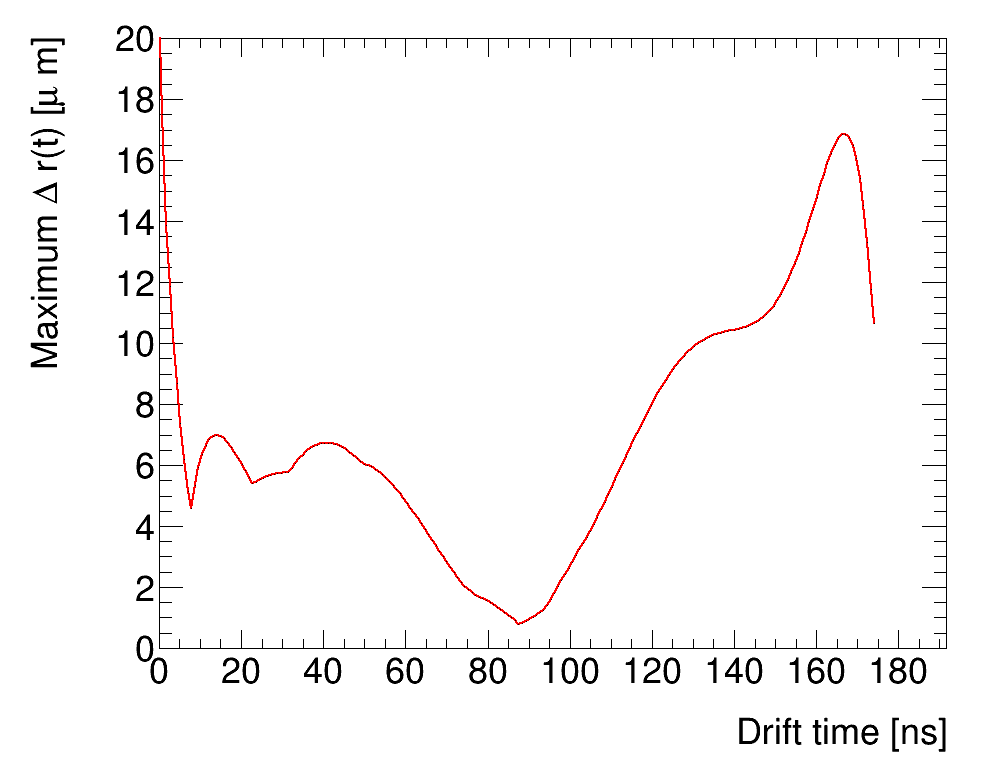}
         \caption{}
         \label{fig:maxRTDiff}
     \end{subfigure}
     \hfill
    \caption{(a) Difference in r(t) function calculated on independent data sets.  Differences are generally on the order of 5-10 microns.  (b) Maximum difference between any two data partitions.  Differences are greatest near the wire and tube wall, but are again on the order of 5-10 microns.}.
    \label{fig:dataPartition}
\end{figure}

The choice of Chebyshev polynomials, and the exact value of the coefficients, is a potential source of systematic uncertainty.  In order to assess the reproducibility of the auto-calibration algorithm we ran the algorithm on independent data sets and compared the resulting r(t) functions.  The results are shown in figure \ref{fig:dataPartition}.  Differences in r(t) functions are on the order of 5-10 microns for the majority of the drift time phase space.  In order to assess the impact on the final resolution result we again calculated the residual distributions (figures \ref{fig:fitResiduals} and \ref{fig:hitResiduals}) but used an r(t) function from a different data partition.  We then applied the usual deconvolution, convolution, and resolution calculation and find the maximum change in resolution is 0.3 \textmu m.  We use this value as the uncertainty estimate for the Chebyshev parameterization and auto-calibration routine.

\subsection{Track Fitting}\label{sec:unc_trackfit}

In order to calculate the uncertainty associated with track fitting we used the likelihood ratio technique.  We varied the track parameters ($\Delta t_0$, angle, and impact parameter) around their mean values and found the values which cause the likelihood ratio to change by one standard deviation.  The size of the perturbation in each parameter required to shift the likelihood by one sigma is shown in table \ref{tab:trackParSigma}.    After perturbing the track parameters, we recreated the biased and unbiased residual distributions and propagated the perturbed residual distributions through the deconvolution procedure to measure the resolution.
From this method we find a systematic error of  5.5 \textmu m due to the track fit.  See table \ref{tab:trackParSigma} for a complete summary of how the individual parameters impact the resolution.


\begin{table}[!htbp]
    \centering
    \begin{tabular}{|c|c|c|} \hline
        Track Parameter & Mean 1 sigma deviation & Change in resolution \\ \hline 
        $\Delta t_0$ & 0.81 ns & 0.4 \textmu m\\
        $b$ & 28 \textmu m & 5.2 \textmu m\\
        $\theta$ & 140 \textmu rad & 5.5 \textmu m \\ \hline
    \end{tabular}
    \caption{Mean size of systematic shift for track parameters and impact on resolution.}
    \label{tab:trackParSigma}
\end{table}

\subsection{Soft-Electron Component of Cosmic-Ray Spectrum}\label{sec:softEsyst}

In this section we will estimate the systematic effects of the MC procedure.  The primary nuisance parameter is the soft-electron component of the cosmic-ray spectrum, which was nominally chosen to be 30\%.  The deconvolution procedure also introduces some systematic uncertainties in the Fourier transform, because the result needs to be passed through a low-pass filter to remove noise.  The noise is the result of dividing two doubles near the limit of precision in some bins. We assess the uncertainty associated with the low-pass cutoff and the cosmic-ray electron component simultaneously by varying the soft-electron fraction to 25 and 35\% of the total spectrum and re-computing the deconvolution and final resolution.  To be especially conservative we even vary the soft-electron component down to 0\%, and we find that the change in single-hit resolution is small.  The maximum variation in resolution observed was 1.7 \textmu m.  While this may seem small given that electrons are low energy and larger scattering, a significant fraction of electrons are so low energy that they do not penetrate the entire chamber and pass reconstruction cuts.

\subsection{Time Slew Correction}
\begin{figure}
    \centering
    \begin{subfigure}[]{0.49\textwidth}
        \centering
        \includegraphics[width=\textwidth]{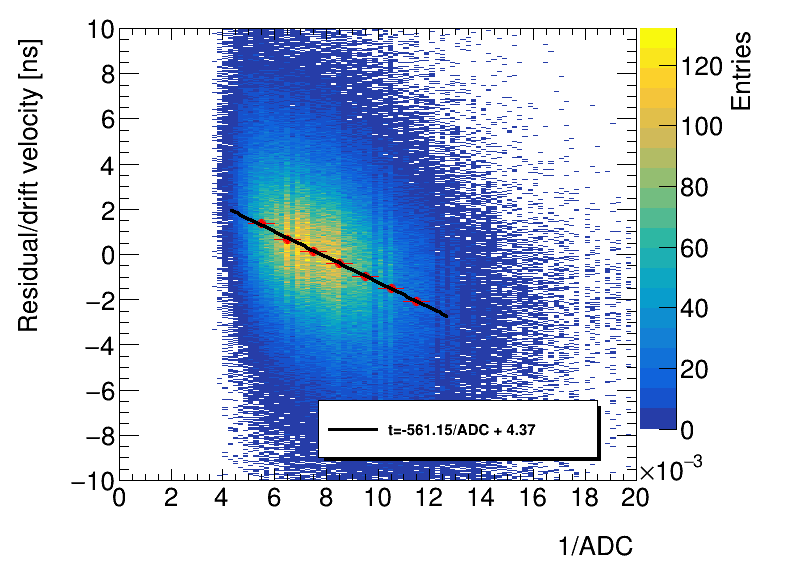}
        \caption{}
        \label{fig:ts_0}
    \end{subfigure}
    \hfill
    \begin{subfigure}[]{0.49\textwidth}
        \centering
        \includegraphics[width=\textwidth]{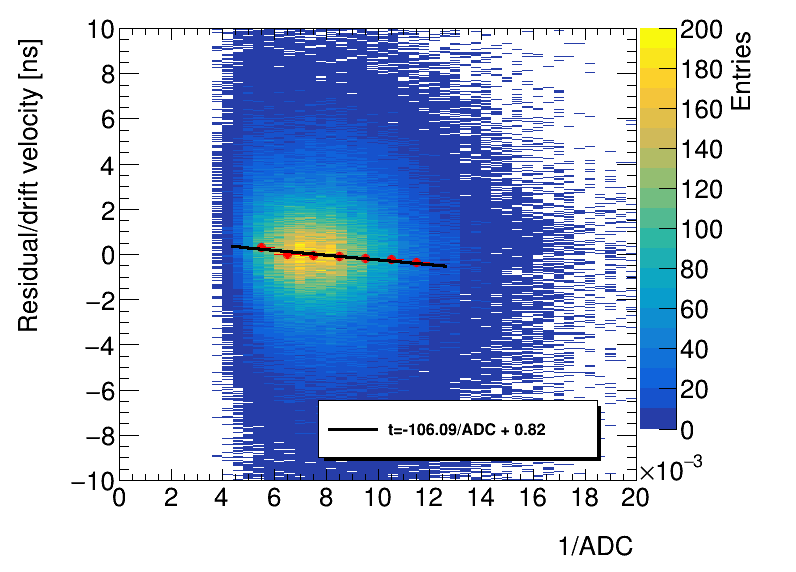}
        \caption{}
        \label{fig:ts_nom}
    \end{subfigure}
    \vskip\baselineskip
    \begin{subfigure}[]{0.49\textwidth}
        \centering
        \includegraphics[width=\textwidth]{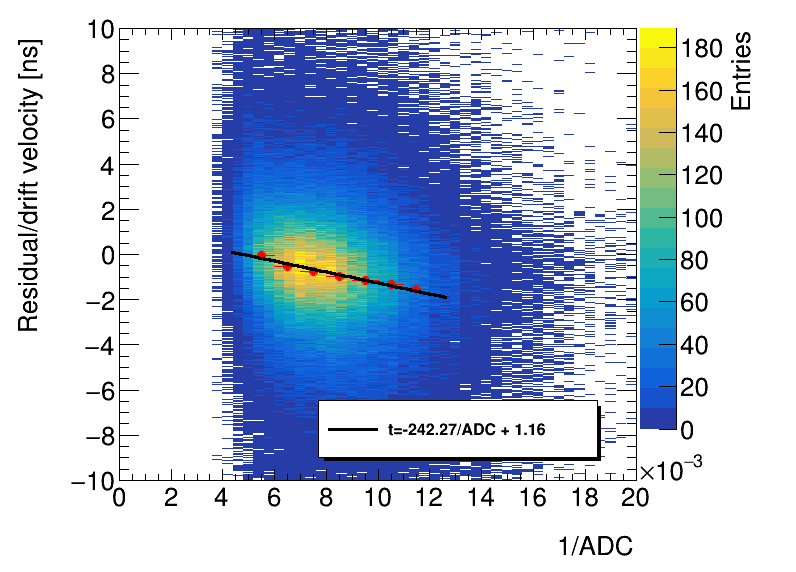}
        \caption{}
        \label{fig:ts_down}
    \end{subfigure}
    \hfill
    \begin{subfigure}[]{0.49\textwidth}
        \centering
        \includegraphics[width=\textwidth]{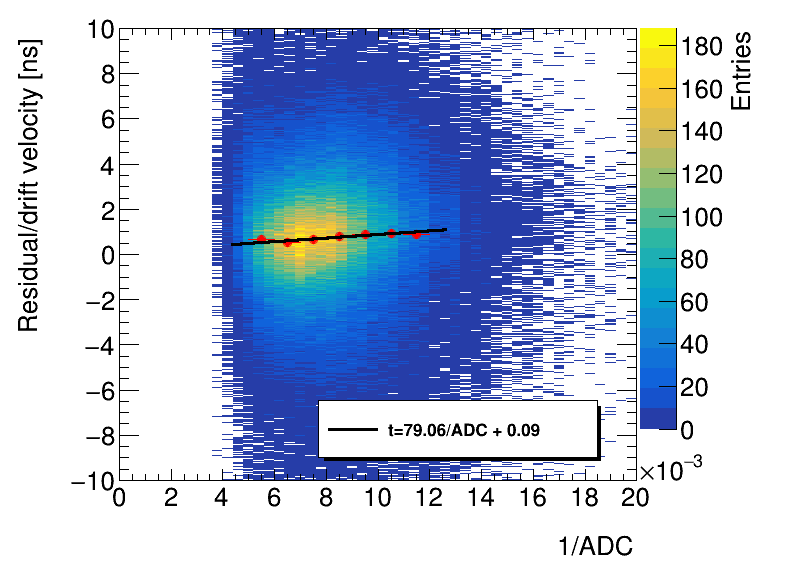}
        \caption{}
        \label{fig:ts_up}
    \end{subfigure}
    \caption{ADC dependence of residuals (a) before and (b) after applying time slew correction, as well as with the amplitude of the time slew correction varied 25\% (c) down and (d) up.  The residuals are divided by the average drift velocity to put the y axis in units of nanoseconds.  The red dots show the mean residual/drift velocity in coarse bins and the black line is a straight line fit to this coarsely binned data.  The text boxes show the equations found for the straight line fits.}
    \label{fig:time_slew_syst}
\end{figure}

The adopted values for the time slew correction function shown in equation (\ref{eq:timeslew}) have uncertainty.  In order to estimate the size of this uncertainty we plot the inverse ADC dependence of the residuals before and after applying the time slew correction in figures \ref{fig:ts_0} and \ref{fig:ts_nom}, respectively.  The ideal time slew correction would have a slope of zero in the straight line fit in figure \ref{fig:ts_nom}.  However, a slight correlation remains after applying our time slew correction.  Figures \ref{fig:ts_down} and \ref{fig:ts_up} show the relationship between residuals and ADC with the amplitude of the time slew correction varied down or up by 25\%, respectively.  The 25\% variation is large enough that the straight line fits plotted in figures \ref{fig:ts_down} and \ref{fig:ts_up} have slopes of opposite sign, showing that our chosen systematic variation is large enough to account for the degree to which our choice of time slew correction amplitude is non-optimal.
Therefore we adopt a conservative upper limit of 25\% as the uncertainty on the amplitude of the time slew correction.

The uncertainty in the amplitude of the time slew correction is propagated to the resolution by taking the maximum difference between the resolution measured using the nominal time slew correction and the two systematic variations, with amplitude varied up and down by 25\%.  The impact on the resolution is 3.7 \textmu m.

\subsection{Signal Propagation Time}
In our test set up we cannot measure the position
of the hit along the tube length, and thus we cannot make an explicit correction for the propagation time of the signal from hit location to readout electronics.   The mean signal propagation time is included in the $\Delta t_0$ parameter, but inclined tracks may have an additional variations of propagation time.

Propagation time is determined by signal propagation speed and distance.
First, we address signal propagation speed.  A drift tube is a coaxial transmission line whose signal propagation speed affected by two factors: (i) the dielectric medium and (ii) the skin depth correction.  The dielectric medium used in this analysis is a 93:7 \% admixture of argon and carbon dioxide at 3 bars absolute pressure.  The permittivity of the gas mixture is similar to a vacuum and so we need only calculate the skin depth correction.  Using the inductance per unit length of a coaxial cable from \cite{Jackson} and assuming a permeability of 1 and a skin depth of 84 microns for 1 MHz signals in aluminum we find a signal propagation speed of 77\% the speed of light.

To estimate the systematic error due variations
in signal propagation distance we do as follows.
We generate tracks with random angle drawn from a from a $\cos^2\theta$ distribution \cite{PDG}, up to the maximum geometric angular acceptance (defined by total chamber height and length of wires).  
We calculate the signal propagation distance for each hit as a function of $\theta$ and subtract the mean propagation distance (as the mean propagation time is already accounted for by the $\Delta t_0$ parameter).  Using the signal propagation speed, we convert the distance to a signal propagation time.  This procedure shows the standard deviation of the signal propagation time is about 0.4 nanoseconds.  Then, assuming hits are uniformly distributed in radius, we choose a random radius between 0 and the inner tube radius, and use the r(t) function 
to calculate the difference in measured radius if the timing was displaced by the signal propagation time.  Hits at small radii are more sensitive to changes in timing because the drift velocity is larger near the wire.  We deconvolute the resulting biased and unbiased residuals with this histogram, and find a change in resolution of 2.2 \textmu m.  Thus, 2.2 \textmu m is the systematic uncertainty envelope associated with the signal propagation timing.

To check the estimate of the signal propagation uncertainty we replaced the original 1.5 m$^2$ scintillator with a smaller one of 0.2 m$^2$.  We observe the difference in residual width between the top layer, closest to the scintillator, and the bottom layer, farthest from the scintillator. The signal propagation time will be much more constrained in the top layer than the bottom layer, leading to a wider residual distribution in the bottom layer.  Since the scintillator is 20 cm long, there is still a small amount of signal propagation smearing in the top layer.  We observe that the resolution in the top layer is 1.9 \textmu m narrower than the resolution measured in the bottom layer.  This is consistent with the 2.2 \textmu m result obtained via theoretical calculations.  Therefore, with empirical evidence and supporting calculations, we conclude that the signal propagation uncertainty is 2.2 \textmu m.

\subsection{Systematic uncertainties on efficiency measurement}

\begin{figure}[!htbp]
    \centering
    \includegraphics[width=0.5\textwidth]{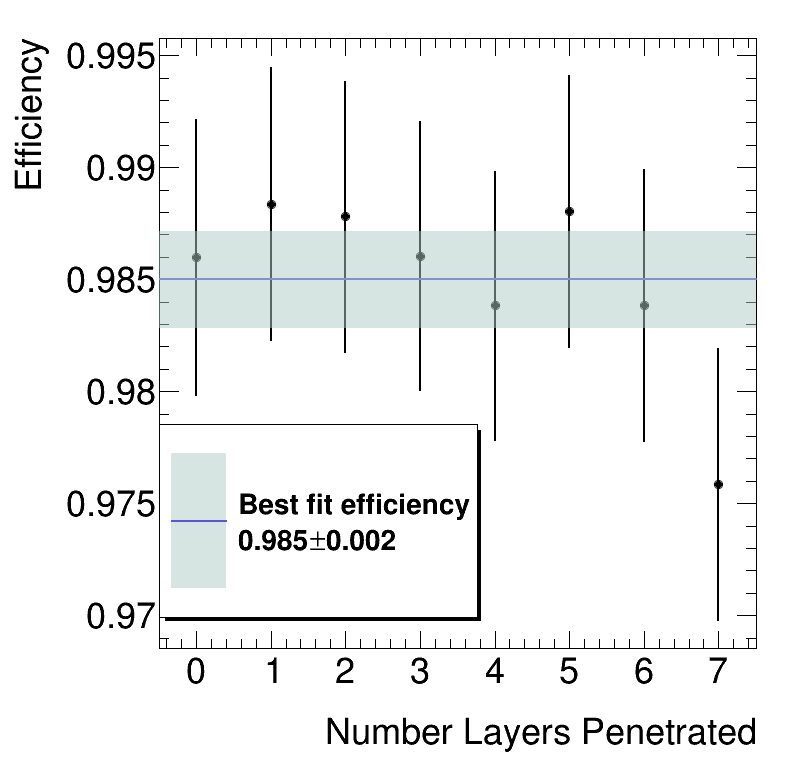}
    \caption{Efficiency within the gas volume measured as a function of the number of layers penetrated.  0 layers penetrated corresponds to the top layer, 1 to the second from the top layer, and so on.  The constant fit has a chi-square per degree of freedom 0.45, and is not in tension with the data.  We do not adopt a systematic uncertainty in this case.}
    \label{fig:layerEff}
\end{figure}

A possible source of systematic bias in the efficiency measurement is the fact that we average the efficiency of all 8 layers.  Cosmic-rays are incident from above.  If there are soft particles (i.e. electrons) that can penetrate some but not all of the layers, it is possible that the efficiency is a function of the number of layers penetrated.  We plot the tube efficiency as a function of the number of layers penetrated in figure \ref{fig:layerEff} (geometric inefficiencies are not a function of layers penetrated).  The constant fit with error bars is also plotted. The constant fit has a chi-square per degree of freedom of 0.45, and is not significantly different than the observed data.  Therefore we do not adopt any systematic uncertainty regarding the efficiency as function of the number of layers penetrated. 

%% file: Results.tex
\subsection{Single-Hit Resolution}

As discussed in section \ref{sec:resolution}, the resolution of the BMG prototype chamber was observed to be 117.7$\pm2.1$ \textmu m (statistical uncertainty only) without accounting for multiple scattering.  When both the biased and unbiased residuals were deconvoluted using the procedure outlined in section \ref{sec:deconvolution} the resolution improved to 103.7$\pm3.5$ \textmu m.  Accounting for the systematic uncertainty described in section \ref{sec:syst} the observed resolution is 103.7$\pm 8.1$ \textmu m.  For the BIS prototype chamber we found the resolution with both ASD-1 and ASD-2 chips.  Notably, the ASD-2 chip achieves finer resolution due to its higher gain. Using ASD-1 chips on the BIS chamber yields a resolution of 101.8$\pm$7.8 \textmu m whereas using ASD-2 chips improves the resolution to 83.4$\pm$7.8 \textmu m.  The ASD-2 chips will be used on sMDT chambers installed in the ATLAS detector, and will be installed on all existing ATLAS MDT chambers in the ATLAS Phase-II upgrade in 2024.  See figure \ref{fig:resVsRadius} for a comparison of the resolution of the BMG and BIS prototype chambers with the Run 2 MDT result for reference \cite{MDTResolution}. 

The systematic uncertainties for both chambers are summarized in table \ref{tab:resolution_systs}.  The systematic uncertainties are different due to the differing geometry of the two chambers and the different readout electronics (ASD-1 vs 2).  While the two chambers are very similar (same operating voltage, tube radius) there are geometric differences that can subtly impact the performance.  Most notably that the two multilayers are approximately 15 cm closer in the BIS prototype as compared to the BMG chamber.  Chambers are designed with different geometries in order to fit in various locations within the ATLAS muon spectrometer.

\begin{table}[]
    \centering
    \begin{tabular}{|l|c|c|} \hline 
        Uncertainty & BMG prototype & BIS prototype \\ \hline
        Statistical & 3.5 \textmu m  & 2.4 \textmu m \\
        Track Fit Parameters Error Propagation & 5.5 \textmu m  & 5.2 \textmu m\\
        Soft-Electron Component in MC simulation & 1.7 \textmu m & 3.0 \textmu m\\
        Signal Propagation Time & 2.2 \textmu m & 3.3 \textmu m \\
        Time Slew Correction Normalization & 3.7 \textmu m & 2.8 \textmu m  \\
        r(t) parameterization & 0.3 \textmu m & 0.3 \textmu m\\ \hline
        Total & 8.1 \textmu m & 7.8 \textmu m \\ \hline
    \end{tabular}
    \caption{Summary of error bars for single-hit resolution measurement.  Total error is obtained by adding the individual errors in quadrature.}
    \label{tab:resolution_systs}
\end{table}

We have also calculated the resolution as a function of radius.  We binned the biased and unbiased residuals for both the MC truth and observed data in 1~mm radial bins and performed the deconvolution procedure in each bin.  The resulting resolution vs radius curve is shown in figure \ref{fig:resVsRadius}.  The resolution vs radius plots have been fit with a second order polynomial whose coefficients are summarized in table \ref{tab:resVsRadiusPolyPars}.  As discussed in section \ref{sec:method}, the resolution procedure is performed iteratively until the coefficients in table \ref{tab:resVsRadiusPolyPars} converge.  The coefficients define the error on a hit, depending on its radius.

\begin{table}[!htbp]
    \centering
    \begin{tabular}{|c|c|c|} \hline
        Parameter & BMG Prototype Value [\textmu m] & BIS Prototype Value [\textmu m]\\ \hline 
        $c_0$ & 243.2$\pm$10.9 & 213.4$\pm$10.6 \\
        $c_1$ & -56.2$\pm$6.5  & -56.6$\pm$6.4\\
        $c_2$ & 4.8$\pm$0.9 & 5.3$\pm$0.8 \\ \hline
    \end{tabular}
    \caption{Fitted polynomial coefficients for the residual vs. radius curve, a function of drift distance.} 
    \label{tab:resVsRadiusPolyPars}
\end{table}

\subsection{Efficiency}
\begin{figure}[!htbp]
\centering
    \begin{subfigure}[]{0.49\textwidth}
        \centering
        \includegraphics[width=\textwidth]{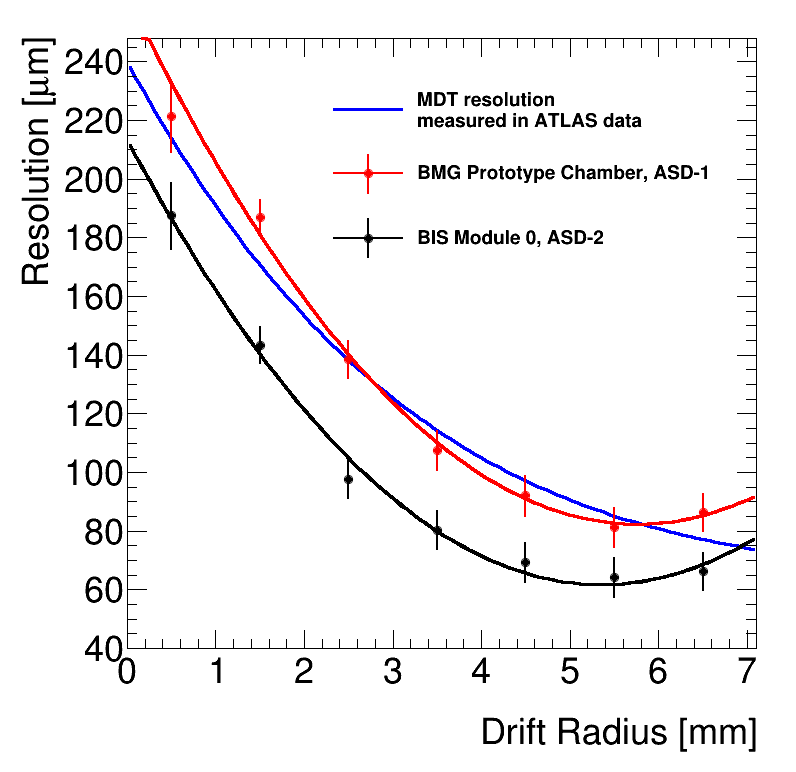}
        \caption{}
        \label{fig:resVsRadius}
    \end{subfigure}
    \hfill
    \begin{subfigure}[]{0.49\textwidth}
        \centering
        \includegraphics[width=\textwidth]{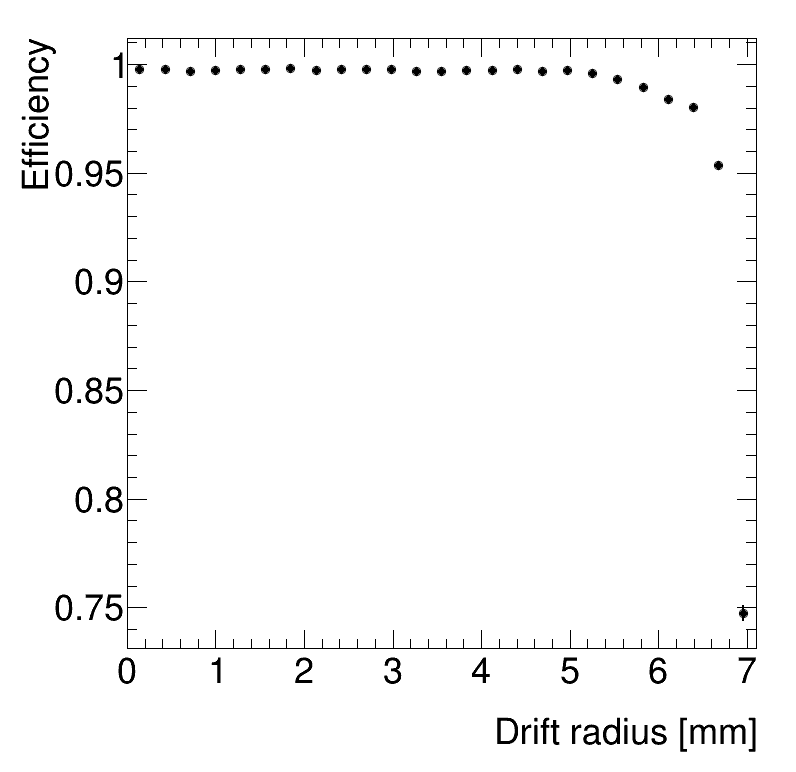}
        \caption{}
        \label{fig:efficiency}
    \end{subfigure}
    \caption{(a) Observed resolution as a function of radius, after performing the deconvolution technique to remove the effects of multiple scattering.  For comparison we show the best fit curve observed in the Run 2 MDT resolution study~\cite{MDTResolution}. (b) Efficiency as a function of drift radius shown with statistical error bars measured on the BMG prototype chamber.  The region near the tube wall is responsible for most of the inefficiency of the chamber within the gas volume.  The 0.9 mm gap between gas volumes, consisting of the two 0.4 mm tube walls and the 0.1 mm space between tubes, is the reason for the rest of the inefficiency.}
    \label{fig:results}
\end{figure}

Using data taken on the BMG prototype sMDT chamber the measured (expected) layer efficiency is 0.942$\pm$0.002 (0.94) \cite{smdtDesign}.  The measured tube efficiency (when only considering the active gas volume of the tubes) is 0.985$\pm$0.002.  figure \ref{fig:efficiency} shows the efficiency as a function of drift radius, which demonstrates that the region near the tube wall is responsible for most of the inefficiency of the chamber, other than the intrinsic geometric inefficiency due to spacing between the gas volumes.  Similar results are achieved with the BIS chamber.

\begin{figure}
\centering
    \begin{subfigure}[]{0.49\textwidth}
        \centering
        \includegraphics[width=\textwidth]{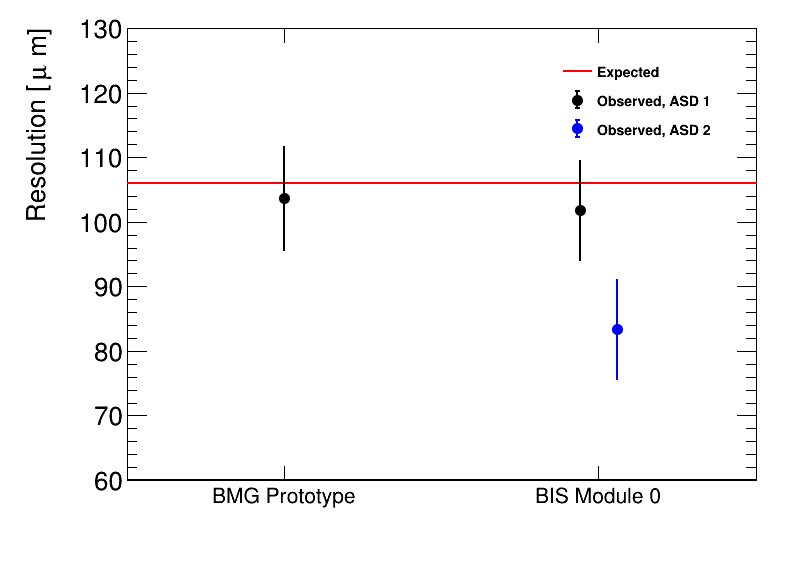}
        \caption{}
        \label{fig:res_summary}
    \end{subfigure}
    \hfill
    \begin{subfigure}[]{0.49\textwidth}
        \centering
        \includegraphics[width=\textwidth]{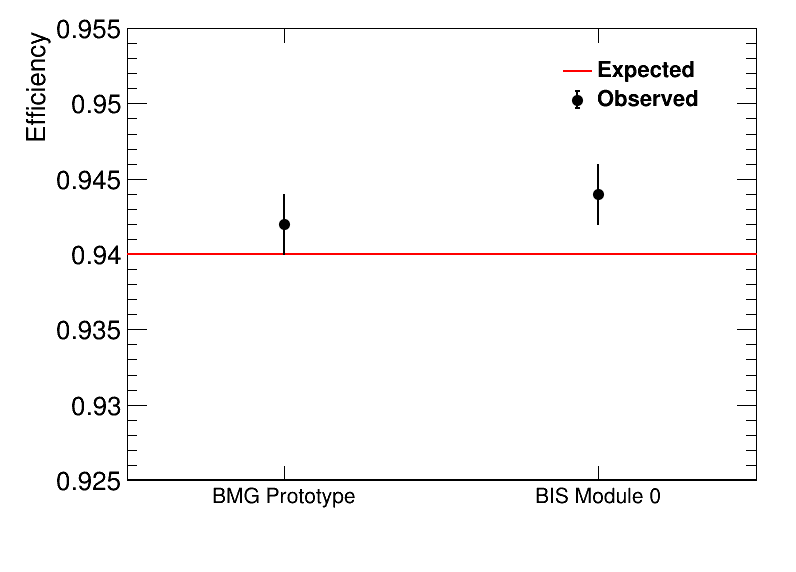}
        \caption{}
        \label{fig:eff_summary}
    \end{subfigure}
    \caption{(a) Measured resolution for the BMG and BIS prototype chambers.  BIS resolution was measured with both legacy electronics (ASD-1) and the new ASD-2 chips, which have higher gain.  (b) Measured efficiency for BMG and BIS chambers}
    \label{fig:summary_plots}
\end{figure}

We did not observe any relationship between efficiency and the number of layers penetrated, so the quoted efficiency is an average of all 8 layers.  See figure \ref{fig:eff_summary} for a comparison of the efficiencies measured on the prototype chambers.

%% file: Conclusions.tex
We observed resolutions of 103.7$\pm8.1$ \textmu m and $101.8\pm7.8$ \textmu m for the BMG and BIS prototype sMDT chambers, respectively. These measurements are consistent with the expected resolution of 106 \textmu m which comes from previous studies of sMDT performance \cite{Schmidt-Sommerfeld:2713422,Hadzic:2692069}.   The BIS chamber, when instrumented with the higher gain ASD-2 chips,
improved to a resolution of $83.4\pm7.8$ \textmu m.  The three crucial components of the analysis strategy were the in-situ r(t) calibration, the addition of a $\Delta t_0$ parameter, and measurement of multiple scattering via Geant simulation.  The r(t) was determined by an auto-calibration using a linearized least squares procedure.
The addition of the $\Delta t_0$ parameter is necessary to account for trigger timing smeared by a large trigger scintillator.  The convolution procedure was motivated by the Monte Carlo simulation, which showed a significant difference in multiple scattering for cosmic-rays and 20 GeV muons.  By accounting for multiple scattering we were able to measure the prototype chamber resolution as it would perform on muons with 20 GeV of $p_T$. The measured efficiency was 0.942$\pm$0.002 (0.944$\pm$0.002) for the BMG (BIS) chamber, consistent with the expected efficiency 0.94. This efficiency is mainly due to the 0.9 mm gap  between neighboring active gas volumes (two 0.4mm tube walls and 0.1mm spacing between drift tubes).  We conclude that the sMDT prototype chambers assembled at the University of Michigan are consistent with the design specifications for resolution and efficiency.

%% file: main.bbl
\providecommand{\href}[2]{#2}\begingroup\raggedright\begin{thebibliography}{10}

\bibitem{AtlasCollaboration_2008}
{\scshape ATLAS} collaboration, \emph{The {ATLAS} experiment at the {CERN}
  large hadron collider},
  \href{http://dx.doi.org/10.1088/1748-0221/3/08/s08003}{\emph{JINST}
  {\bfseries 3} (2008) S08003}.

\bibitem{TDRMuonSpectrometer}
{\scshape ATLAS} collaboration, \emph{{ATLAS Muon Spectrometer : Technical
  Design Report}},   CERN-LHCC-97–022, https://cds.cern.ch/record/331068.

\bibitem{phase2TDRTDAQ}
{\scshape ATLAS} collaboration, \emph{{Technical Design Report for the Phase-II
  Upgrade of the ATLAS TDAQ System}},   CERN-LHCC-2017-020,
  https://cds.cern.ch/record/2285584.

\bibitem{smdtDesign}
H.~Kroha, R.~Fakhrutdinov and A.~Kozhin, \emph{{New High-Precision Drift-Tube
  Detectors for the ATLAS Muon Spectrometer}},
  \href{http://dx.doi.org/10.1088/1748-0221/12/06/C06007}{\emph{JINST}
  {\bfseries 12} 12}.

\bibitem{TDCandMiniDAQ}
Y.~Guo, J.~Wang, Y.~Liang, X.~Xiao, X.~Hu, Q.~An et~al., \emph{{Design of a
  Time-to-Digital Converter ASIC and a mini-DAQ system for the Phase-2 upgrade
  of the ATLAS Monitored Drift Tube detector}},
  \href{http://dx.doi.org/10.1016/j.nima.2020.164896}{\emph{Nucl. Instrum.
  Meth. A} {\bfseries 988} (2021) 164896}.

\bibitem{asdManual}
{\scshape ATLAS} collaboration, C.~Posch, E.~Hazen and J.~Oliver,
  \emph{{MDT-ASD}, {CMOS} front-end for {ATLAS} {MDT}},   ATL-MUON-2002-003,
  https://cds.cern.ch/record/684217.

\bibitem{HPTDC}
J.~Christiansen, \emph{{HPTDC High Performance Time to Digital Converter}},
  https://cds.cern.ch/record/1067476, CERN, 2004.

\bibitem{TTCvi}
P.~Farthouat and P.~Gällnö, \emph{{TTC-VMEbus INTERFACE}},
  https://ttc.web.cern.ch/TTCviSpec.pdf.

\bibitem{MDTResolution}
{\scshape ATLAS} collaboration, G.~Aad et~al., \emph{{Resolution of the ATLAS
  muon spectrometer monitored drift tubes in LHC Run 2}},
  \href{http://dx.doi.org/10.1088/1748-0221/14/09/P09011}{\emph{JINST}
  {\bfseries 14} (2019) P09011}.

\bibitem{autocalibration}
C.~Bacci, C.~Bini, G.~Ciapetti, G.~{De Zorzi}, P.~Gauzzi, F.~Lacava et~al.,
  \emph{Autocalibration of high precision drift tubes},
  \href{http://dx.doi.org/10.1016/S0920-5632(97)00128-X}{\emph{Nuclear Physics
  B - Proceedings Supplements} {\bfseries 54} (1997) 311--316}.

\bibitem{Carnegie}
R.~Carnegie, M.~Dixit, J.~Dubeau, D.~Karlen, J.-P.Martin, H.~Mes et~al.,
  \emph{{Resolution studies of cosmic ray tracks in a TPC with GEM readout}},
  \href{http://dx.doi.org/10.1016/j.nima.2004.08.132}{\emph{Nucl. Instrum.
  Meth. A} {\bfseries 538} (2005) 372--383}.

\bibitem{Geant4}
S.~Agostinelli, J.~Allison, K.~Amako, J.~Apostolakis, H.~Araujo, P.~Arce
  et~al., \emph{Geant4-a simulation toolkit},
  \href{http://dx.doi.org/10.1016/S0168-9002(03)01368-8}{\emph{Nucl. Instrum.
  Meth. A} {\bfseries 506} (2003) 250 -- 303}.

\bibitem{PDG}
{\scshape Particle Data Group} collaboration, \emph{Review of particle
  physics}, \href{http://dx.doi.org/10.1103/PhysRevD.98.030001}{\emph{Phys.
  Rev. D} {\bfseries 98} (2018) 030001}.

\bibitem{Jackson}
J.~D. Jackson, \emph{{Classical electrodynamics; 2nd ed.}}
\newblock Wiley, New York, NY, 1975.

\bibitem{Schmidt-Sommerfeld:2713422}
{\scshape ATLAS} collaboration, K.~R. Schmidt-Sommerfeld, \emph{{Study of Muon
  Drift Tube Detectors and Fast Readout Electronics for Very High Counting
  Rates}},   CERN-THESIS-2020-016, https://cds.cern.ch/record/2713422.

\bibitem{Hadzic:2692069}
{\scshape ATLAS} collaboration, S.~Hadzic, \emph{{Test of High-Resolution Muon
  Drift-Tube Chambers for the Upgrade of the ATLAS Experiment}},
  CERN-THESIS-2019-150, https://cds.cern.ch/record/2692069.

\end{thebibliography}\endgroup
